\documentclass[twocolumn,astrosymb,tighten,trackchanges,twocolappendix]{aastex701}

\usepackage{physics}
\usepackage{natbib}
\usepackage{mathtools}
\usepackage{graphicx}	                
\usepackage{amsmath}	                
\usepackage{bm}                         
\usepackage{enumitem}
\usepackage{url}
\usepackage{braket}
\usepackage{threeparttable}
\usepackage{placeins}

\usepackage[final]{zebra-goodies}     
\zebranewnote{question}{orange}
\zebranewnote{revised}{red}

\newcommand\dex{\,{\rm dex}}

\newcommand\mpc{\, h^{-1}{\rm {Mpc}}}

\newcommand\Myr{\,{\rm Myr}}
\newcommand\Gyr{\,{\rm Gyr}}

\newcommand\kms{\,{\rm {km\, s^{-1}}}}
\newcommand\Msun{\,{\rm M_\odot}}
\newcommand\msun{\, h^{-1}{\rm M_\odot}}
\newcommand\msunperyr{\, {\rm M_\odot}{\rm yr}^{-1}}

\newcommand\Lsun{\, {\rm L_\odot}}
\newcommand\perccm{\, {\rm cm}^{-3}}

\newcommand\Hatom{{\rm H}}
\newcommand\Hmol{{{\rm H}_2}}

\newcommand\figem{\bf}                  
\newcommand\textem{\em}                 
\def\softwarenamestyle[#1]{\textsc{#1}}

\defcitealias{moTwophaseModelGalaxy2024}{Paper-I}
\defcitealias{chenTwophaseModelGalaxy2024}{Paper-II}
\defcitealias{chenTwophaseModelGalaxy2025}{Paper-III}
\defcitealias{chenTwophaseModelGalaxy2025a}{Paper-IV}

\shorttitle{LRDs as BHs in super-Eddington accretion}
\shortauthors{Chen and Mo}

\begin{document}

\title{Super-Eddington accretion of black holes in early nuclear bursts gives birth to Little Red Dots}

\author[orcid=0000-0002-4597-5798,gname=Yangyao,sname=Chen]{Yangyao Chen}
\affiliation{School of Astronomy and Space Science, Nanjing University, Nanjing, Jiangsu 210093, China}
\affiliation{Key Laboratory of Modern Astronomy and Astrophysics, Nanjing University, Ministry of Education, Nanjing, Jiangsu 210093, China}
\email[show]{yangyaochen.astro@foxmail.com}  

\author[orcid=0000-0001-5356-2419,gname=Houjun,sname=Mo]{Houjun Mo} 
\affiliation{Department of Astronomy, University of Massachusetts, Amherst, MA 01003, USA}
\email[show]{hjmo@umass.edu}


\begin{abstract}
In a recent paper, Chen et al. developed a framework for modeling the seeding 
and growth of supermassive black holes (BHs) in the context of $\Lambda$CDM cosmogony.   
Here, we use a set of physically motivated criteria to select a population of predicted BHs 
and link them to Little Red Dots (LRDs) discovered by JWST. We show that the 
LRD population at high redshift ($z$) emerges naturally from a subset of BHs 
with super-Eddington accretion during nuclear bursts. 
The model suggests that the observed LRDs are the ``tip of the iceberg''
of a much larger population of less luminous BHs in the same subset.
The model makes specific predictions for the LRD population, such as the mass distributions
of their BHs and host galaxies/halos, and the piece-wise redshift evolution of their number density.
The cosmological context of the model also allows us to link the 
observed LRD population to their progenitors (their BH seeds) and lower-$z$ 
descendant BHs, galaxies and halos. 
Most LRDs at $z\sim 5$ are seeded at 
$z \gtrsim 20$ through direct-collapse BHs or pair-instability supernovae from 
Pop-III stars, and have grown to $M_{\rm BH} \approx 10^5$--$10^7\Msun$ through 
nuclear bursts by their observed redshift. 
LRDs are predicted to have diverse descendants, ranging from compact dwarf galaxies
to brightest cluster galaxies (BCGs) at $z=0$. These predictions 
are consistent with current observations and can be further tested.
The success of this model indicates that the results presented here
provide a robust foundation for building detailed models of the LRD population.
\end{abstract}



\section{Introduction}
\label{sec:intro}


%
%


%

Recent JWST observations unveiled a population of Little Red Dots (LRDs)
at high redshift ($z$), characterized by compact morphology, a peculiar V-shape spectral 
energy distribution (SED), the presence of broad emission lines,
and a number density much higher than that of known quasars \citep[e.g.][]{mattheeLittleRedDots2024,kokorevCensusPhotometricallySelected2024,greeneUNCOVERSpectroscopyConfirms2024,perez-gonzalezWhatNatureLittle2024,akinsCOSMOSWebOverabundancePhysical2025,labbeUNCOVERCandidateRed2025,
chenHostGalaxyIf2025,liLittleRedDots2025}.
Follow-up multi-wavelength observations showed that LRDs are deficient
in X-ray, infrared, submillimeter and radio emission
\citep{yueStackingXRayObservations2024,pergerDeepSilenceRadio2025,akinsCOSMOSWebOverabundancePhysical2025,greeneWhatYouSee2026}.
These observed properties of LRDs are distinct from known populations 
of active galactic nuclei (AGNs) and galaxies, making their existence 
an interesting problem for galaxy formation and black hole (BH) growth 
in the $\Lambda$CDM cosmogony.

Much effort has been put into understanding the nature and origin of LRDs.
Hypotheses that have been proposed and examined include 
changing the properties of dust to accommodate the peculiar shape of 
the observed SED \citep{liLittleRedDots2025}; 
invoking scattered light from the AGN and young stellar population to 
explain the rest-frame UV part of the SED \citep{greeneUNCOVERSpectroscopyConfirms2024,
labbeUNCOVERCandidateRed2025,inayoshiSpectralUniformityLittle2026}; 
associating the central BH with a stellar-like envelope \citep{degraaffRemarkableRubyAbsorption2025,kidoBlackHoleEnvelopes2025,naiduBlackHoleStar2025,inayoshiSpectralUniformityLittle2026};
linking LRDs to the formation of BH seeds \citep{begelmanLittleRedDots2026,baggenConnectingDotsUVbright2026} 
to produce the rest-frame red part of the SED;
assuming the gas column density to be in the Compton-thick regime to produce the lack of 
X-ray emission \citep{liuBalmerBreakOptical2025,jiBlackTHUNDERNonstellarBalmer2025,naiduBlackHoleStar2025};
and assuming self-interacting dark matter to transport heat outward so as to produce 
collapsed central objects in dark matter halos \citep{jiangFormationLittleRed2025,wangHaloAssemblyBias2026,robertsLittleRedDots2026}. 
These efforts provide important insights into the parameter space that
could be relevant to the formation of LRDs. However, a self-consistent 
and coherent scenario in the context of the current structure-formation paradigm 
is still lacking. 

The main challenge in modeling the formation of LRDs from first principles 
comes from the complex interplay among processes operating in the high-$z$ universe 
and the large dynamic ranges involved. For example, the 
interpretation based on ongoing seeding events invokes strong heating 
to suppress the fragmentation of gas and star formation prior to the observed epochs 
of LRDs (e.g. \citealt{baggenConnectingDotsUVbright2026}; see also 
\citealt{greifDelayPopulationIII2011};
\citealt{wiseFormationMassiveBlack2019};
\citealt{latifTurbulentColdFlows2022} for alternative heating mechanisms). 
However, at $z \approx 5$ when LRDs are abundant, previous star formation may
already have enriched the intergalactic medium (IGM; e.g. Fig.~D4 
of \citetalias{chenTwophaseModelGalaxy2025a}; \citealt{spinosoMultiflavourSMBHSeeding2023}),
which makes the effectiveness of such heating questionable. 
The solution with a gas envelope and nuclear star formation surrounding the 
BH may be able to reproduce the observed SED \citep[e.g.][]{degraaffRemarkableRubyAbsorption2025,kidoBlackHoleEnvelopes2025,naiduBlackHoleStar2025,inayoshiSpectralUniformityLittle2026}, but complex baryon cycles, associated with the expected bursty star formation 
for small galaxies at high $z$ 
\citep{el-badryBreathingFIREHow2016,sternVirializationInnerCGM2021,hopkinsWhatCausesFormation2023,asadaBurstyStarFormation2024,maTwophaseFormationGalaxies2026}, make the depletion and replenishment of 
gas highly uncertain. Modeling these processes using current cosmological hydrodynamical 
simulations suffers from uncertainties in sub-grid implementation, such as artificial 
seeding of BHs and simplified treatment of BH accretion and feedback \citep[e.g.][]{sijackiUnifiedModelAGN2007,weinbergerSimulatingGalaxyFormation2017,koudmaniUnifiedAccretionDisc2024,liPhysicalProcessesCoevolution2025}. Idealized and zoom-in simulations 
\citep[e.g.][]{suSelfregulationBlackHole2023,hopkinsFORGEdFIREResolving2024,sivasankaranAGNFeedbackIsolated2025} 
may miss the cosmological context and lack the power to make statistical predictions.  

Recently, \citet[hereafter \citetalias{chenTwophaseModelGalaxy2025a}]{chenTwophaseModelGalaxy2025a} 
constructed a physical framework to seed and grow BHs together with their host 
galaxies and halos within the $\Lambda$CDM cosmogony.
This model is based on the two-phase scenario of galaxy formation proposed by 
\citet[hereafter \citetalias{moTwophaseModelGalaxy2024}]{moTwophaseModelGalaxy2024},
with key extensions needed to account for the seeding and early growth of BHs.
The model tackles the aforementioned challenge by self-consistently modeling 
the complex interplay among processes relevant to BH seeding and growth,
and uses semi-analytical approximations with multi-layer numerical 
refinements to achieve the resolution needed to cover the full structure 
hierarchy from large-scale structures to the surroundings of BHs.
As shown in \citet[hereafter \citetalias{chenTwophaseModelGalaxy2024}]{chenTwophaseModelGalaxy2024}, 
\citet[hereafter \citetalias{chenTwophaseModelGalaxy2025}]{chenTwophaseModelGalaxy2025} 
and \citetalias{chenTwophaseModelGalaxy2025a},
the model not only reproduces a wide range of observed properties of BHs and their 
host galaxies/halos, but also predicts the evolution histories of BHs from 
their seeding times to $z=0$. It is therefore interesting to 
study whether the model can yield a population of objects that resemble the observed LRDs. 
In particular, the cosmological context of the model allows us to 
investigate not only the physical mechanisms that give birth to LRDs 
but also their fates and links to other objects at lower-$z$.
As we show, our model predicts a branch of BHs that are in 
super-Eddington accretion during nuclear bursts.   
A subset of these BHs closely resemble the observed LRDs 
in accretion rates and in BH mass to stellar mass ratios, 
and the model can be used to predict their number density, 
their host galaxy/halo and descendant properties.

This paper is organized as follows. In \S\ref{sec:method}, we summarize the 
model of galaxy formation and BH growth, and we define the selection criteria 
for LRDs. In \S\ref{sec:predictions}, we present model predictions for LRDs, 
their host galaxies/halos, and their descendants. In \S\ref{sec:summary}, we summarize our results
and discuss their implications.

\section{The method}
\label{sec:method}


Here we first provide a brief description of the model of galaxy formation and BH growth 
developed in \citetalias{chenTwophaseModelGalaxy2025a}.
We then define criteria to select LRDs from the predicted 
properties of BHs and host galaxies.

\begin{figure} \centering
    \includegraphics[width=\columnwidth]{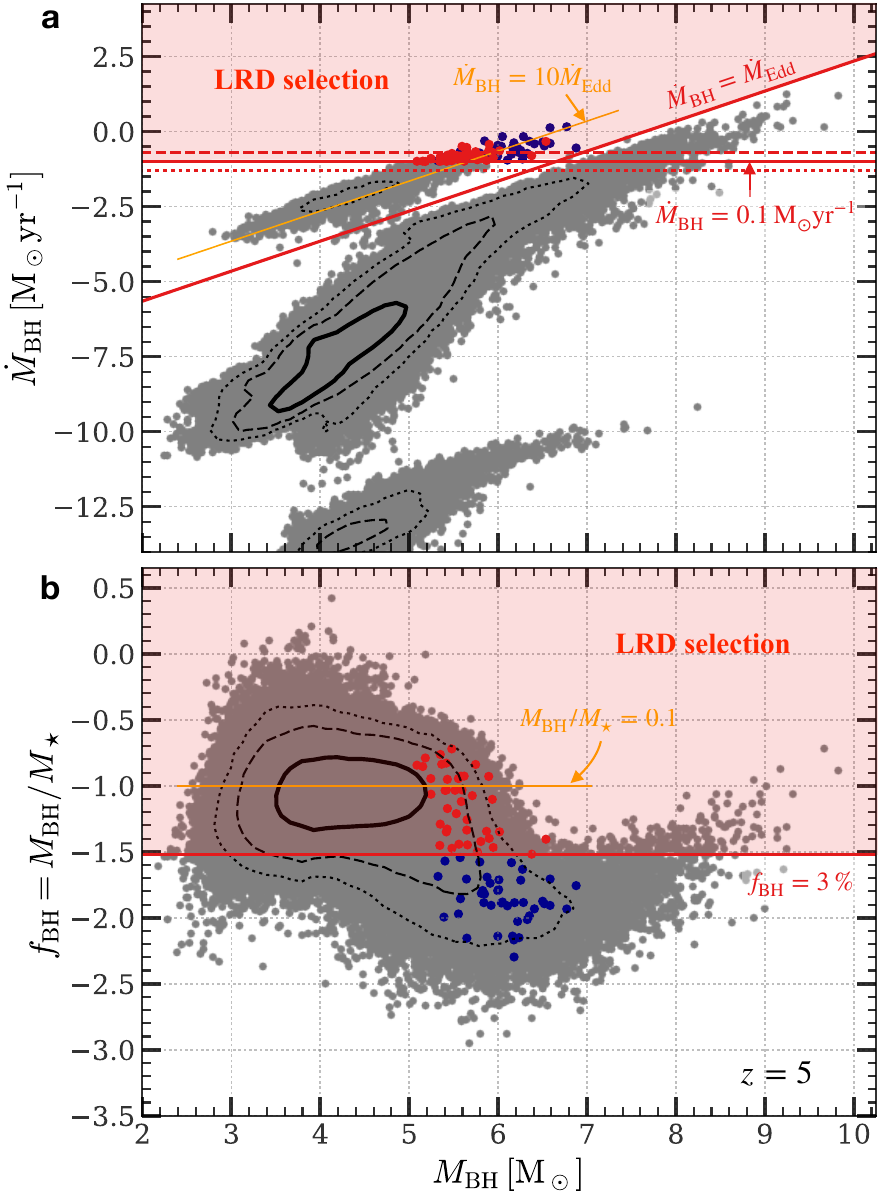}
    \caption{
    {\figem Selection of LRDs.}
    Each dot represents a BH produced by the model at $z=5$ in the 
    planes of {\figem a}, BHAR ($\dot{M}_{\rm BH}$) versus BH mass, 
    and {\figem b}, BH-mass fraction ($f_{\rm BH} \equiv M_{\rm BH}/M_\star$)
    versus BH mass. 
    In each panel, {\figem contours} from inner to outer encompass 
    $68\%$, $95\%$ and $99.7\%$, respectively, of all BHs. 
    {\figem Orange line} indicates the typical $\dot{M}_{\rm BH}$
    or $f_{\rm BH}$ of the bursty branch.
    The default selection criterion for LRDs is marked by {\figem red shading} 
    enclosed by {\figem red solid lines}, and LRDs so selected are shown by 
    {\figem red dots}. BHs passing the selection for $\dot{M}_{\rm BH}$
    but not for $f_{\rm BH}$ are shown by {\figem blue dots}.
    {\figem Red dashed and dotted lines} in {\figem a} show alternative selection 
    criteria with varied thresholds of BHAR.
    The figure shows that the observed LRDs are the tip of the branch
    shaped by nuclear bursts (see \S\ref{ssec:sample} for details).
    }
    \label{fig:selection}
\end{figure}

\subsection{The model of galaxy formation and black hole growth}
\label{ssec:model}


The model takes halo merger trees from TNG100-1-Dark \citep{pillepichFirstResultsIllustrisTNG2018, nelsonIllustrisTNGSimulationsPublic2019}. The simulation, the same as 
used in \citetalias{chenTwophaseModelGalaxy2025a}, has a periodic box 
of $75\mpc$ on a side and a particle mass of $6.0\times 10^6 \msun$. 
The cosmology adopted in the simulation is a flat $\Lambda$CDM cosmology
\citep{planckcollaborationPlanck2015Results2016}, with parameters of 
$h=0.6774$, $\Omega_{\rm m,0}=0.3089$, $\Omega_{\Lambda,0}=0.6911$, 
$\Omega_{\rm b,0}=0.0486$, and a Gaussian initial power
spectrum specified by $n_{\rm s}=0.9667$ and $\sigma_8=0.8159$.
The same cosmology is assumed in our analysis. 

The sample of simulated subhalos used here is selected to have 
$M_{\rm halo} \geq 10^{9.5}\Msun$ at $z=0$ for a central subhalo and at the time 
of infall for a satellite subhalo. The assembly history of each 
subhalo is constructed directly from the simulation above a mass 
limit of $5\times 10^8 \msun$. An extension algorithm is applied to trace the assembly history of each halo 
further back to a mass of $M_{\rm halo} \approx 10^5\Msun$ (see Appendix D1 of \citetalias{chenTwophaseModelGalaxy2025a}; 
see also \citealt{chenELUCIDVICosmic2019,chenConditionalAbundanceMatching2023}), 
so that 
processes leading to the formation of the first-generation stars and BH seeds 
in the halo can be modeled. This eliminates the need for artificially assigning 
massive BH seeds to halos 
\citep[e.g.][]{somervilleSemianalyticModelCoevolution2008,crainEAGLESimulationsGalaxy2015,weinbergerSimulatingGalaxyFormation2017},
which is crucial as LRDs may be powered by BHs of relatively low mass.   
The use of simulated halos also allows us to model the environments
of individual halos. 
 
A seeding procedure is applied to each assembly history 
to identify the epoch of first collapse of the halo gas and 
to determine the properties of first-generation stars and BH seeds 
formed in the collapsed gas (see \S2.3 of \citetalias{chenTwophaseModelGalaxy2025a}). 
The key components of the seeding procedure can be summarized as follows.
\begin{enumerate}[parsep=0pt,itemsep=0pt,leftmargin=0pt]
     \item 
     An environmentally modulated criterion for gas collapse. Here
     internal coolants ($\Hatom$ and $\Hmol$) are evaluated
     to identify the halo mass $M_{\rm halo}$ at which gas in a pristine halo 
     can effectively cool and collapse. 
     Three environmental effects that can advance or delay the collapse 
     are included: the Lyman-Werner (LW) radiation produced
     by other galaxies, which can dissociate $\Hmol$ and thus delay the 
     collapse; the dynamical heating due to fast halo accretion
     in the early Universe, which can heat the halo gas and thus 
     also delay the collapse; the enrichment of the IGM, which can feed 
     coolants into the halo and thus can advance the collapse.
     \item A feedback-regulated initial mass function (IMF). Here, the feedback 
     from the first Pop-III star cluster formed in the collapsed 
     gas of a pristine halo is evaluated to determine the resulting mass of individual
     stars in the cluster.
     \item A `multi-flavor' stellar evolution. Here, the evolution of individual
     stars in the first cluster is followed according to their initial 
     masses. The mass of the most massive star in the cluster determines
     whether a core-collapse supernova (CCSN), a pair-instability supernova (PISN), 
     or a direct collapse black hole (DCBH) can occur.
     The most massive BH remnant in the cluster is then defined as the seed
     of the central BH,
     while other stars constitute the initial stellar component of the host galaxy.
\end{enumerate}
Thus, for each halo, the output of the seeding procedure is the initial $M_\star$ of the galaxy, 
the $M_{\rm BH}$ of the BH seed, as well as the flavor (DCBH, PISN or CCSN)
that represents the evolution pathway of the most massive star in the first 
star cluster. 

A growing procedure then starts with the initial conditions set by the seeding 
procedure to model the post-seeding growth of the BHs and their host galaxies
(see \S2.4 of \citetalias{chenTwophaseModelGalaxy2025a}).
The key feature here is the multi-channel growth of BHs, 
with each channel representing a way to generate dynamical hotness, 
to remove the angular-momentum barrier, and to feed baryons to the central BH.
These channels can be summarized as follows.
\begin{enumerate}[parsep=0pt,itemsep=0pt,leftmargin=0pt]
    \item The bursty mode, which appears as rapid BH growth via 
    super-Eddington accretion associated with intense starbursts 
    in the dense gaseous nucleus formed under a global disturbance 
    of the host galaxy.
    The gas density of such a nucleus can reach a `supernova-free' 
    threshold, $n_{\rm snf} \equiv 10^{3.5}\perccm$,
    corresponding to a dynamical timescale of $t_{\rm dyn, snf} = 0.75 \Myr$
    that is too short for the feedback of massive stars to be important. 
    The consequence of the formation of such a nucleus is a ``nuclear burst''
    characterized by super-Eddington accretion of the central BH,
    intense star formation that grows the nuclear star cluster (NSC),
    and strong BH feedback that disperses the nucleus.
    The trigger of a nuclear burst is modeled once the host halo 
    temporarily experiences an excursion of fast assembly that brings in 
    a large amount of matter so as to disturb the gravitational potential of the 
    host galaxy significantly.
    The duration of a nuclear burst is $\lesssim t_{\rm dyn, snf}$ 
    before the mode is quenched and another excursion is triggered.
    \item The continuous mode, in which the BH continuously captures 
    sub-clouds seeded and amplified by the turbulence 
    in the self-gravitating gas cloud (SGC; see \citetalias{moTwophaseModelGalaxy2024} for the concept) formed 
    during the fast assembly phase of a halo.
    The associated star formation in the SGC is to build an 
    extended, dynamically hot stellar component (bulge) of the host 
    galaxy. In contrast, the collapse of the gas in the slow phase of a halo is expected 
    to form a dynamically cold disk in which the growth of the BH ceases.
    \item The merger mode, in which two galaxies and their central BHs 
    merge to form a new galaxy and a new central BH.
\end{enumerate}
These channels are not mutually exclusive at a given time during the post-seeding 
growth of a BH. However, the dominating channel transits sequentially 
from the bursty mode, via the continuous mode, to the merger mode over the lifetime 
of a BH (see \S3.2 of \citetalias{chenTwophaseModelGalaxy2025a}).
As will be shown in \S\ref{ssec:n_vs_z}, the transition of the dominant channel 
is the key to explaining the observed evolution in the number density of LRDs 
over cosmic time. The product of the growing procedure is the properties of BHs and their 
host galaxies/halos, as well as the contribution of each growth channel to the 
total growth, from the seeding epoch to $z=0$.

Since the duration of a nuclear burst is much shorter than the simulation 
snapshot interval, we assume that it occurs over a random duration 
$[t,t+t_{\rm dyn,snf}]$, where $t$ is sampled uniformly from the
time interval spanned by the snapshot in which the trigger is identified.
The black hole accretion rate (BHAR, denoted as $\dot{M}_{\rm BH}$) 
during the burst is then calculated as the ratio 
between $\Delta M_{\rm BH}$ and $t_{\rm dyn,snf}$.

Fig.~\ref{fig:selection}a shows the distribution of individual BHs predicted by 
the model at $z=5$ in the $\dot{M}_{\rm BH}$-$M_{\rm BH}$ plane.
A prominent multi-modal distribution can be seen. The dominant branch of the 
distribution, which is defined by the continuous mode and contains most of the 
BHs in our sample (see the contours), extends from $M_{\rm BH} \sim 10^2\Msun$ to $10^9\Msun$ 
and exhibits a positive correlation between $\dot{M}_{\rm BH}$ and $M_{\rm BH}$. 
The lower branch represents the passive population in which BH growth is halted 
either as the host galaxy becomes a satellite or the host halo temporarily
makes an excursion to slow assembly. Of particular interest is the upper branch, 
which represents the population of nuclear bursts and is characterized by 
$\dot{M}_{\rm BH}\sim 10\times {\dot M}_{\rm Edd}$, with the Eddington accretion 
rate defined as  
\begin{multline}
    \dot{M}_{\rm Edd} \equiv \frac{M_{\rm BH}}{\epsilon_{\rm r}t_{\rm Sal}}
    = 2.22 \times 10^{-2} \times \\
    \left(\frac{M_{\rm BH}}{10^6\Msun}\right) \left(\frac{\epsilon_{\rm r}}{0.1}\right)^{-1}
    \msunperyr
    \,.
\end{multline}
Here, $t_{\rm Sal} = 0.450\Gyr$ is the Salpeter timescale, 
and $\epsilon_{\rm r}\sim 0.1$ is the radiative efficiency 
\citep[e.g.][]{yuanHotAccretionFlows2014,weinbergerSimulatingGalaxyFormation2017}.

Fig.~\ref{fig:selection}b shows the distribution of BHs in the 
$f_{\rm BH}$-$M_{\rm BH}$ plane, where BH-mass fraction
$f_{\rm BH} \equiv M_{\rm BH}/M_\star$ is defined as the ratio of BH mass to stellar mass 
of the host galaxy.
The multi-channel growth of BHs can also be seen in this distribution. 
The frequent triggers of nuclear bursts in the early Universe can lead to a 
high BH-mass fraction of $f_{\rm BH} \approx 0.1$ -- a consequence
of the supernova-free nature of nuclear bursts and ineffective BH feedback 
due to photo-trapping at high accretion rate. The scatter 
of $f_{\rm BH}$ in this regime is large, due to the stochastic nature of 
the triggers and the rapid growth in individual bursts. In contrast,
the continuous mode can sustain a lower $f_{\rm BH}$ with 
a smaller scatter due to the continuous capturing of sub-clouds 
and the regulation of BH growth by both stellar and BH feedback.
The dominant channel of the BH growth makes a transition from the bursty mode 
to the continuous mode at $M_{\rm BH} \sim 10^6$--$10^7\Msun$, 
above which BHs are massive enough to generate strong feedback 
to effectively quench the bursty mode.

\subsection{Little red dots as black holes in super-Eddington accretion}
\label{ssec:sample}

We use the following set of criteria to select LRDs from the predicted BH population
at any given $z$.
\begin{enumerate}[parsep=0pt,itemsep=0pt,leftmargin=0pt]
    \item 
    The BH must accrete at a super-Eddington rate, 
    $\dot{M}_{\rm BH} \geq \dot{M}_{\rm Edd}$. 
    This criterion is motivated by theoretical considerations \citep[e.g.][]{degraaffRemarkableRubyAbsorption2025,kidoBlackHoleEnvelopes2025,naiduBlackHoleStar2025,liuBalmerBreakOptical2025,inayoshiSpectralUniformityLittle2026,trincaYouCantSee2026} that such an accretion is able to produce 
    a gas envelope around the BH to explain the observed SED of LRDs.
    This selection also separates the LRD candidates from the 
    dominant sub-Eddington branch in Fig.~\ref{fig:selection}a, 
    making the selected sample a distinct population 
    rather than an extension of other BHs in the 
    $\dot{M}_{\rm BH}$-$M_{\rm BH}$ plane.
    \item 
    The BH has a sufficiently high BHAR, 
    $\dot{M}_{\rm BH} \geqslant \dot{M}_{\rm BH,min}$,
    so that it is luminous enough to be observed and identified as an LRD.
    \item 
    The BH-mass fraction is sufficiently high, $f_{\rm BH} \equiv M_{\rm BH}/M_\star \geq f_{\rm BH,min}$,
    so that the stellar component of the host galaxy is not too bright to obscure the 
    spectral features of the BH. 
\end{enumerate}
The value of $\dot{M}_{\rm BH,min}$ depends on the detection limits 
of current observations. Here we take $\dot{M}_{\rm BH,min} = 0.1\msunperyr$ as 
the default to match the observed number density of LRDs at $z \gtrsim 4$ 
(see \S\ref{ssec:n_vs_z}). We also use alternative criteria, with
$\dot{M}_{\rm BH,min} = 0.05\msunperyr$ and $0.2\msunperyr$ 
to demonstrate the dependence on this threshold (see \S\ref{sec:predictions}).
These choices are also consistent with some other evaluations in the 
literature \citep[e.g.][]{begelmanLittleRedDots2026,inayoshiSpectralUniformityLittle2026}. 

The other threshold, $f_{\rm BH,min}$, can be estimated using  
the condition that the total radiation from the BH dominates over 
that of the stellar component of the host galaxy so that AGN features 
are not overwhelmed by stellar lights. As suggested by the model 
implemented here (see Appendix C5 of \citetalias{chenTwophaseModelGalaxy2025a}), 
the bolometric luminosity ($L_{\rm BH}$) of a BH in super-Eddington accretion
is limited to around the Eddington level, 
\begin{equation}
    L_{\rm BH} \approx L_{\rm Edd} = 
    3.28 \times 10^4 \Lsun \left(\frac{M_{\rm BH}}{\rm M_\odot}\right)
    \,.
\end{equation}
We estimate the bolometric luminosity of a stellar population younger than 
a few Myr using {\sc FSPS} \citep{conroyPropagationUncertaintiesStellar2009,
conroyPropagationUncertaintiesStellar2010}, as
\begin{equation}
    L_\star \approx 10^3\Lsun\left(\frac{M_\star}{\rm M_\odot}\right)
    \,,
\end{equation}
where a Chabrier IMF \citep{chabrierGalacticStellarSubstellar2003} 
and a solar metallicity are assumed. The requirement of 
$L_{\rm BH} \geqslant L_\star$ then leads to a condition of 
$M_{\rm BH}/M_\star \geqslant 0.0304$. We thus set our default threshold of 
$f_{\rm BH,min} = 3\%$. 
We note that the exact value of $f_{\rm BH,min}$ is uncertain, as it depends on 
the exact SEDs of both the BH and the stellar components, 
the radiation transfer, and the selection of LRDs in observations
\citep[e.g.][]{marszewskiLittleRedDots2026,shenLuminaProjectDemographics2026,
greeneUNCOVERSpectroscopyConfirms2024}. 
Our luminosity-based argument thus only provides a crude estimate, and we will discuss the effects of varying this threshold in \S\ref{sec:predictions}.

Fig.~\ref{fig:selection} shows the boundaries of the default selection
criterion (red solid lines) and regions enclosed by these boundaries (red shading)
predicted at $z = 5$. BHs passing only the first two conditions are highlighted in blue, 
while those also passing the third one are highlighted in red.
Alternative criteria with varied thresholds of $\dot{M}_{\rm BH}$ are 
also indicated (red dashed and dotted lines).
The final sample at $z = 5$ based on the default criterion contains 41 LRDs
in the simulation box, allowing a reliable analysis of the population.
All modeled LRDs fall in the massive tip of the bursty branch in the 
$\dot{M}_{\rm BH}$-$M_{\rm BH}$ plane. {\textem This pinpoints the 
bursty mode as the physical origin of LRDs, and indicates 
that the observed LRD population is the tip of an iceberg whose 
main body is hidden below the waterline set by the current detection limit.}
This also explains why LRDs are not identified prior to JWST, as  
the tip level can only be reached with the JWST sensitivity.
This is also consistent with the fact that no truncation is detected  
in the luminosity function of LRDs at the faint end \citep[e.g.][]{greeneUNCOVERSpectroscopyConfirms2024,kokorevCensusPhotometricallySelected2024,labbeUNCOVERCandidateRed2025}.
In Fig.~\ref{fig:lf}a of Appendix~\ref{app:sec:lf}, we show 
the predicted bolometric
luminosity functions for LRD populations selected by different criteria, and demonstrate that the observed LRDs, regardless of the exact selection 
criterion, are indeed the bright tip of 
a much larger population of BHs in super-Eddington accretion.

As shown in the following, our selection criterion is 
capable of reproducing key properties of the LRD population, 
such as the sharp drop of the number density at $z \lesssim 4$, 
the compact morphology, the blue UV part of the V-shape SED, and the 
stable UV-to-optical luminosity ratio, without fine-tuning 
the model. More importantly, our model treats the formation of LRDs 
within the same cosmological context as the formation of 
other populations of galaxies and BHs, and can make predictions 
that can be tested by observations.

\section{Model predictions}
\label{sec:predictions}

\begin{figure} \centering
    \includegraphics[width=\columnwidth]{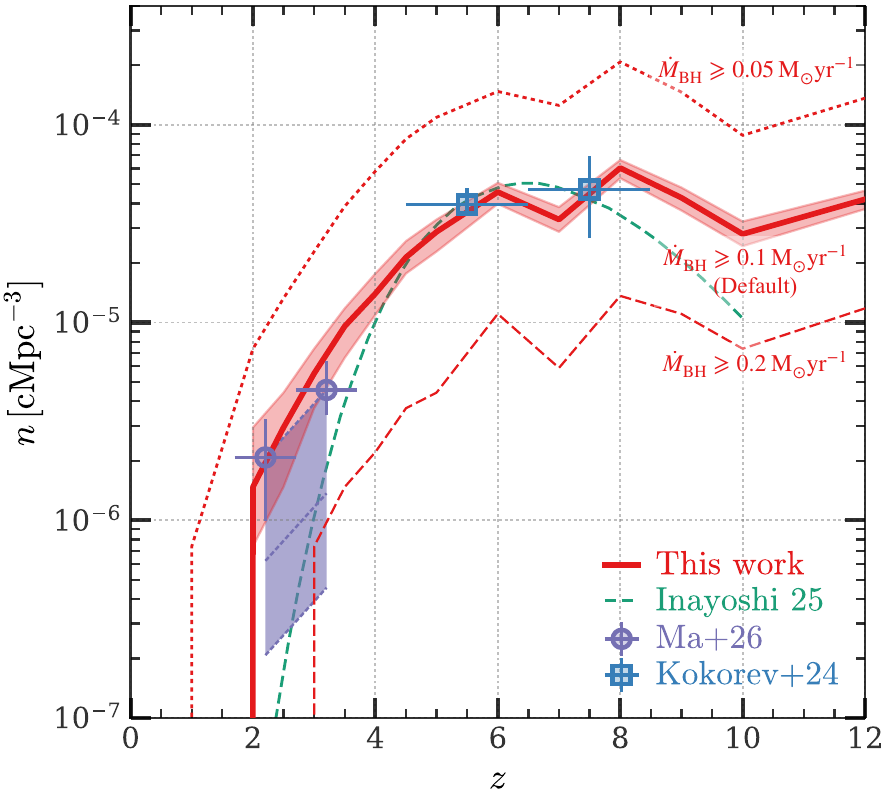}
    \caption{
    {\figem Number density of LRDs as a function of redshift.}
    Here we show the predicted number density ($n$) of LRDs 
    selected by our default criterion ({\figem red solid}) 
    and the criteria with alternative thresholds of 
    $\dot{M}_{\rm BH}$ ({\figem red dashed and dotted}).
    The curve with shading represents the median and $16^{\rm th}$--$84^{\rm th}$ 
    percentile range of $n$ among 100 random samples for the 
    durations of nuclear bursts (see \S\ref{ssec:model}).
    For comparison, we show the observations by \citet{maCountingLittleRed2026}
    and \citet{kokorevCensusPhotometricallySelected2024},
    and empirical result of \citet{inayoshiLittleRedDots2025}.
    Three dotted lines associated with the {\figem purple band}
    indicate the expected number density
    if $100\%$, $30\%$ and $10\%$, respectively, 
    of the selected candidates in \citet{maCountingLittleRed2026} 
    are LRDs. 
    The model recovers the piece-wise evolution of LRD number density
    found in observations,
    and ties the physical origin of this evolution to the hierarchical 
    nature of halo assembly that is determined by the
    $\Lambda$CDM cosmology (see \S\ref{ssec:n_vs_z} for details).
    }
    \label{fig:n_vs_z}
\end{figure}


\subsection{The evolution of number density}
\label{ssec:n_vs_z}

Fig.~\ref{fig:n_vs_z} shows the predicted number density ($n$) of LRDs as a 
function of $z$ based on our default selection (solid red curve).
For comparison, we show the observational results from 
\citet{maCountingLittleRed2026} and \citet{kokorevCensusPhotometricallySelected2024},
and the empirical prediction by \citet{inayoshiLittleRedDots2025}.
The number density of LRDs predicted by our model recovers the piece-wise 
evolution found in observations: a plateau at $z \gtrsim 4$ where the number density reaches a few
times $10^{-5}\,{\rm cMpc}^{-3}$ and a sharp drop at $z \lesssim 4$. 
In addition, the predicted bolometric luminosity function of LRDs with our 
default selection evolves very little in shape at $z \gtrsim 4$, as  
shown in Fig.~\ref{fig:lf}b of Appendix~\ref{app:sec:lf}, 
suggesting that these early epochs are ``stable'' hubs for the formation of LRDs.

The transition of BH growth and star formation from the bursty mode to the 
continuous mode (see \S\ref{ssec:model} and Fig.~\ref{fig:selection}b)
provides a natural explanation for the observed drop of the number density 
of LRDs at $z \lesssim 4$. In the $\Lambda$CDM paradigm, halos with lower 
$M_{\rm halo}$ at lower $z$ are expected to have slower assembly
\citep{dongUniversalSpecificMerger2022,jiangSelfsimilarDecompositionHierarchical2025},
so that their excursions to nuclear bursts are rarer and star formation via the 
continuous mode becomes more important. Consequently, halos that satisfy the three 
conditions for LRDs are increasingly rarer at lower $z$. 
{\textem Our model thus ties the evolution of LRD number density to the 
halo assembly that is determined by the $\Lambda$CDM cosmology}.

Fig.~\ref{fig:n_vs_z} also shows the predicted number density of 
LRDs based on alternative selection criteria with 
$\dot{M}_{\rm BH}$ doubled or halved (red dashed and dotted curves).
There is a factor of $\sim 3$ change in the predicted number 
density. This implies that 
$\dd{n}/\dd{\dot{M}_{\rm BH}} \propto \dot{M}_{\rm BH}^{-2.5}$ 
at $\dot{M}_{\rm BH} \sim 0.1\msunperyr$ for the BH growth driven by the bursty mode,
and further supports the conclusion reached in \S\ref{ssec:sample} 
that a large population of LRDs is still hidden below the current 
detection limit. 
Such a steep decline of $n$ with $\dot{M}_{\rm BH}$ also implies 
a steep bolometric luminosity function of LRDs, as shown in 
Fig.~\ref{fig:lf} of Appendix~\ref{app:sec:lf}.

\begin{figure*} \centering
    \includegraphics[width=0.95\textwidth]{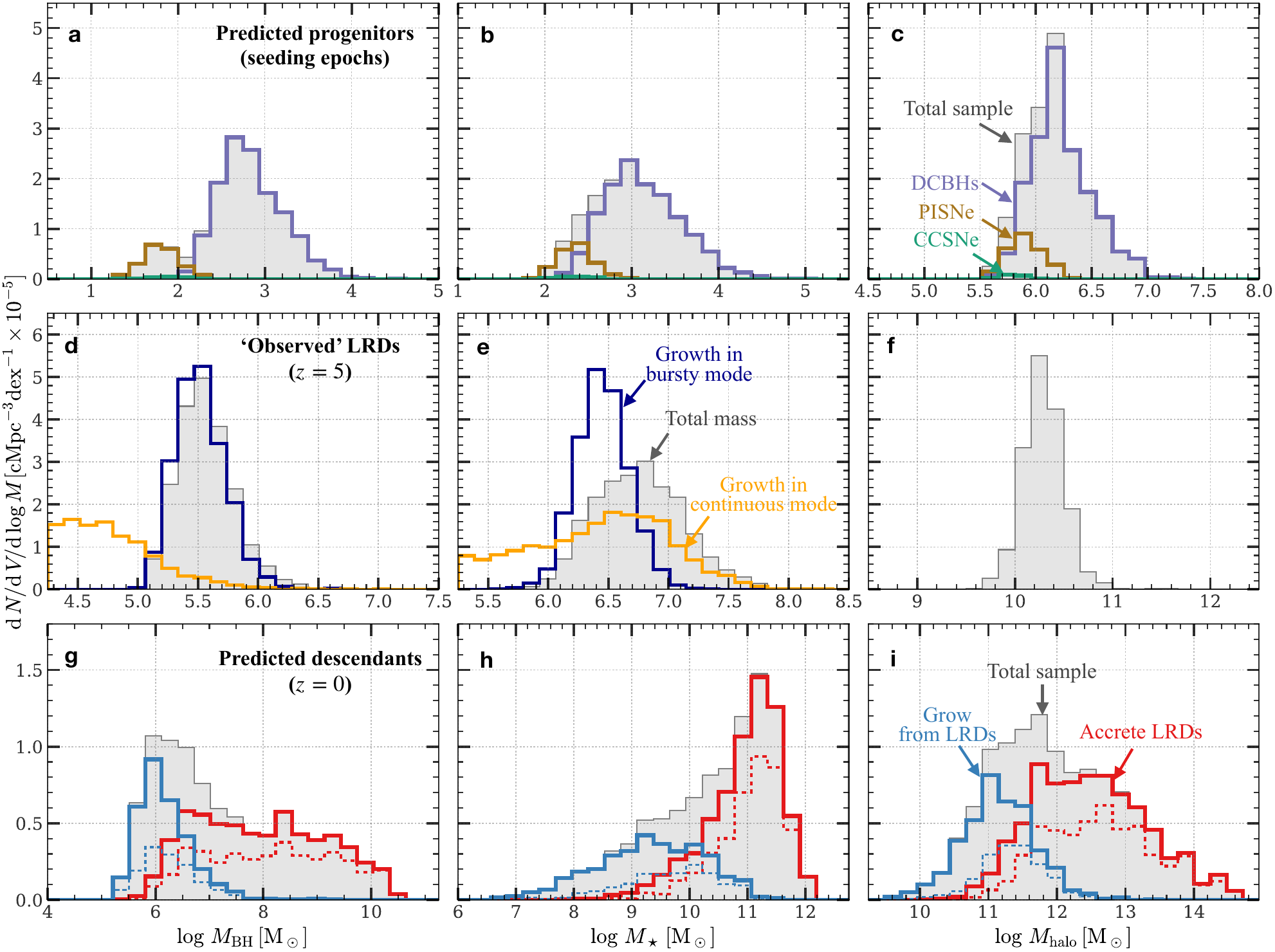}
    \caption{
    {\figem Distribution of properties of LRDs, their progenitors and descendants.} 
    Here the LRDs are those observable at a time slice of $z=5$
    selected by the default criterion 
    (see \S\ref{ssec:sample} and Fig.~\ref{fig:selection}),
    and their progenitors and descendants are identified through their 
    merger trees.
    100 random samples for the durations of nuclear bursts (see \S\ref{ssec:sample}) 
    are drawn and averaged to suppress the sampling noise. 
    The properties shown are $M_{\rm BH}$ of the BH, $M_\star$ of the host galaxy and $M_{\rm halo}$ of the host halo. For satellites, $M_{\rm halo}$ takes the infall value. 
    The distributions are defined as the number of objects per unit 
    comoving volume per unit logarithmic interval of mass,
    and are shown at three stages of their evolution:
    {\figem a}--{\figem c}, the seeding epochs (mostly $z \gtrsim 20$; 
    identified backward in time through the main branch of merger tree);
    {\figem d}--{\figem f}, the ``observed'' epoch ($z=5$); and 
    {\figem g}--{\figem i}, the present day ($z=0$;
    identified forward in time through descendant chain of merger tree).
    At seeding, sub-populations of seeds formed via DCBH, PISN, or CCSN are 
    shown in total or separately.
    At $z=5$, we show the distributions of the total mass, and of those contributed 
    by the bursty and continuous modes separately,
    obtained by accumulating the growth of each object in each 
    mode during all snapshots until $z=5$ 
    (for example, all previous nuclear bursts are counted in the contribution of the 
    bursty mode to an object identified at $z=5$).
    At $z=0$, we show the total sample of descendants, 
    and sub-populations according to how an LRD is connected to its 
    descendant: one includes descendants that grew from LRDs (i.e., LRDs lie on their main branches); the other includes descendants that accreted LRDs through mergers (i.e., LRDs lie on their side branches).
    Within each sub-population at $z=0$, the subset of descendants that are central galaxies is also indicated ({\figem dashed}).
    These archaeological and futurological views pinpoint the venues 
    for the progenitors and descendants of LRDs to be found
    in observations (see \S\ref{ssec:prop_dist} for details).
    }
    \label{fig:prop_dist}
\end{figure*}

\subsection{The progenitors and descendants of LRDs}
\label{ssec:prop_dist}

The complete coverage of our model from the seeding epoch to $z=0$ allows us to 
predict the properties of LRDs, their progenitors at earlier epochs, and their 
descendants at later epochs, in a self-consistent cosmological context.
Fig.~\ref{fig:prop_dist} shows the predicted distributions of 
LRDs observable at a time slice of $z=5$ selected by the default criterion 
at three stages of evolution:
\begin{enumerate}[parsep=0pt,itemsep=0pt,leftmargin=0pt]
    \item 
    the seeding stage, at which the BH seeds of LRDs are formed
    as a result of the first collapse of gas in their host halos
    (see \S\ref{ssec:model} for the seeding procedure); 
    \item 
    the ``observed'' stage, taking $z=5$ as the representative,
    at which BHs are selected as candidates of the observed LRD population;
    \item 
    the evolved stage, taking $z=0$ as an example, at which LRDs have evolved into 
    descendants to be observed in the local Universe. 
\end{enumerate}
At each stage, we show the distributions of $M_{\rm BH}$ of LRDs,
$M_\star$ of their host galaxies, and $M_{\rm halo}$ of their host halos.

At seeding epochs, the BHs of LRD progenitors span a wide range of 
$M_{\rm BH}$, from $\sim 10\Msun$ to $\sim 10^4\Msun$ (Fig.~\ref{fig:prop_dist}a, the gray histogram), 
indicating that diverse environmental processes (see \S\ref{ssec:model}) 
are operating to delay the gas cooling in the ``mini-halos'' 
and to elevate the mass of the first-generation stars formed in the 
collapsed gas. Although not shown here, the seeds of most LRDs are predicted to 
form at $z \gtrsim 20$.
The early formation of these seeds, together with the predicted late occurrence 
of the IGM enrichment (see Fig.~D4 of \citetalias{chenTwophaseModelGalaxy2025a}),
implies that the seeds of LRDs are almost exclusively products of Pop-III stars.
The top-heavy IMF assumed for Pop-III stars in the seeding procedure also 
implies that the initial $M_\star$ of the host galaxy is only a few 
times $M_{\rm BH}$ (Fig.~\ref{fig:prop_dist}b).
Star formation efficiencies in the seeded mini-halos are as low as 
$M_\star/M_{\rm halo} \sim 10^{-3}$ (Fig.~\ref{fig:prop_dist}c), 
due to the dispersal of gas by the feedback from the first star cluster 
in the shallow potential well of such a mini-halo.

The stellar evolution implemented in the seeding procedure 
allows us to divide the seeds of LRDs into sub-populations of different 
flavors (formation pathways). The colored histograms in Fig.~\ref{fig:prop_dist}a 
show the distributions of seeds with CCSN, PISN and DCBH flavors, respectively.
Most seeds are born as DCBHs from (super)massive progenitor stars with masses 
up to $\sim 10^4\Msun$.
A significant fraction of seeds are born as remnants of secondary stars 
after the most massive one in the first cluster is destroyed by a 
PISN. This produces a significant mass gap and bimodality 
in the overall distribution of $M_{\rm BH}$.
Only a small fraction of seeds are born as remnants of CCSN 
from low-mass progenitor stars, indicating that the 
origin of LRDs is quite different from that of stellar-mass BHs 
observed in the local Universe.
The distribution of seed mass $M_{\rm BH}$ falls in the 
intermediate-mass BH (IMBHs) regime, which disfavors the interpretation 
that LRDs are powered by BHs formed directly from the first collapse 
in their host halos \citep[e.g.][]{baggenConnectingDotsUVbright2026}. 
Instead, our model suggests that {\textem it is post-seeding growth, 
mainly through episodic nuclear bursts, that raises BH seeds to 
supermassive status}. We note that the high masses of DCBH seeds
are a prediction of our model, rather than a condition assumed 
to allow these BHs to reach the masses required to power LRDs. 

At $z=5$, the LRD population in our default selection have BHs with 
$M_{\rm BH} \sim 10^5$--$10^6\Msun$ (gray histogram in Fig.~\ref{fig:prop_dist}d).
Decomposing the total $M_{\rm BH}$ into contributions from the bursty 
and continuous modes (colored histograms in Fig.~\ref{fig:prop_dist}d) 
reveals that the bursty mode dominates the BH growth.
The $M_{\rm halo}$ of the host halos are about $\sim 10^{10}$--$10^{11}\Msun$ 
(Fig.~\ref{fig:prop_dist}f), comparable to those of dwarf galaxies in the 
local Universe \citep[e.g.][]{zhangUnexpectedClusteringPattern2025}.
This predicted range of $M_{\rm halo}$ justifies our choice of simulation and 
halo sample to represent potential hosts of LRDs. The low $M_{\rm halo}$ 
implies a relatively weak clustering of LRDs in space, a conclusion that is 
consistent with recent observations \citep{carranza-escuderoLonelyLittleRed2025}.

The $M_\star$ of LRDs is predicted to be $\sim 10^6$--$10^{7.5}\Msun$ 
(Fig.~\ref{fig:prop_dist}e), roughly an order of magnitude higher than 
$M_{\rm BH}$, consistent with the high BH-mass fraction,  
$f_{\rm BH} \sim 0.1$, expected from the growth driven by the bursty mode 
(see Fig.~\ref{fig:selection}b). 
Indeed, decomposing the total $M_\star$ into contributions from the bursty 
and continuous modes (colored histograms in Fig.~\ref{fig:prop_dist}e) 
shows that the bursty mode contributes significantly to the total $M_\star$ 
of LRDs, while the contribution of the continuous mode is only 
marginally higher than that of the bursty mode.
Our model thus suggests that LRDs are transitional objects in 
which the continuous mode of star formation is about to take over the burst mode,
and that the extended stellar components are about to be built 
up in the host galaxies. The dominance of nuclear bursts in the growth of 
LRDs may thus provide an explanation for their observed compact morphology.

Young stellar populations formed via nuclear bursts in an LRD are expected 
to produce a blue UV component in the SED \citep[e.g.][]{greeneUNCOVERSpectroscopyConfirms2024,
labbeUNCOVERCandidateRed2025}. This, together with the red optical component produced by 
the BH in super-Eddington accretion, offers a compelling explanation for the observed 
V-shape SED of LRDs. In addition, the stable $f_{\rm BH} \sim 0.1$ produced by the bursty mode
(see Fig.~\ref{fig:selection}b) implies a near-constant optical-to-UV
luminosity ratio ($L_{\rm opt}/L_{\rm UV}$). Such a constant 
luminosity ratio is found in recent observations, and the value of 
$f_{\rm BH} \sim 0.1$ predicted by our model also aligns with those 
inferred from the SEDs of LRDs 
\citep[e.g.][]{inayoshiSpectralUniformityLittle2026}.

The descendants of LRDs at $z=0$ show a broad distribution of $M_{\rm BH}$
(Fig.~\ref{fig:prop_dist}g), reflecting the diverse evolutionary pathways 
of LRDs after the observed epoch. 
Some LRDs merge into massive galaxies covering a wide range in 
$M_{\rm BH}$, $M_\star$, and $M_{\rm halo}$, from dwarf galaxies to 
the brightest cluster galaxies (BCGs) in massive clusters 
(red solid in Fig.~\ref{fig:prop_dist}g--i). 
Other LRDs evolve along more isolated pathways, growing as the main progenitors of 
their descendants and dominating the lower ends of the mass distributions 
(blue solid in Fig.~\ref{fig:prop_dist}g--i). 
The typical $M_{\rm BH}$ of these low-mass descendants is $\sim 10^6\Msun$,
only a factor of three higher than the typical $M_{\rm BH}$ of LRDs at $z=5$,
implying that their BHs have experienced little growth since $z=5$
and their NSCs are predicted to contain a significant fraction of 
ancient stars formed via nuclear bursts at high $z$.
However, the total $M_\star$ of these low-mass descendants is $\sim 2$--$3\dex$ 
higher than that of the LRDs at $z=5$, implying that extended 
stellar components have been built up in the descendants.
About half of the low-mass descendants remain isolated (blue dashed 
in Fig.~\ref{fig:prop_dist}g--i); the other half are satellites,
susceptible to tidal forces that may strip their extended stellar masses.
This implies a diverse, environment-driven morphology for the descendants, 
ranging from nucleated dwarfs with extended stellar components, 
if they remain isolated, to compact objects such as ultra-compact dwarfs 
(UCDs) or globular-like star clusters, if most of the extended stellar 
masses have been stripped away.

We note that the predicted properties and evolution of LRDs depend
on the selection criterion, which itself varies between observational 
studies. In addition to the dependence of the number density on the luminosity 
threshold controlled by $\dot{M}_{\rm BH,min}$ 
(see \S\ref{ssec:n_vs_z} and Fig.~\ref{fig:n_vs_z}),
the threshold $f_{\rm BH,min}$ also has a noticeable impact on the 
prediction. For example, removing the $f_{\rm BH}$ threshold leads to a 
larger number of LRDs, higher $M_{\rm BH}$ and $M_\star$, and a higher 
fraction of $M_\star$ in their extended stellar components.
The $M_{\rm BH}$ of LRDs may thus reach $\sim 10^7\Msun$ if those having 
larger stellar components are included in the 
sample (e.g. blue points in Fig.~\ref{fig:selection}).
Despite such uncertainties in sample selections, our predictions
are broadly consistent with observations: the predicted 
$M_{\rm BH} \approx 10^5$--$10^7\Msun$ 
for LRDs aligns with those reported by \citet{greeneWhatYouSee2026}
and \citet{lambridesCaseSuperEddingtonAccretion2026}; 
the existence of young, extended stellar components in the host 
galaxies aligns with recent stacking analysis that 
reveals extended, wavelength-dependent emission around LRDs
\citep{zhangUnveilingExtendedComponents2025}; the co-existence 
of NSCs and extended stellar components suggests a bimodal origin 
of the rest-frame UV emission, which may explain the observed 
dichotomy in the distribution of UV half-light radii of LRDs 
\citep{cloonanPANORAMICUVopticalMorphologies2026}. 

\section{Summary and Discussion}
\label{sec:summary}


In this paper, we have applied the model of galaxy formation developed 
in \citetalias{chenTwophaseModelGalaxy2025a} to study the physical origin of LRDs, to 
interpret their observed properties, and to make predictions for their progenitors 
and descendants over cosmic time.

Our key finding is that LRDs 
emerge as a ``tip-of-iceberg'' population of BHs in super-Eddington accretion, 
associated with the ongoing formation of compact NSCs
and with extended stellar components that are about to be built 
in their host galaxies (\S\ref{sec:method}; Fig.~\ref{fig:selection}).
The emergence of this population is driven by nuclear bursts triggered 
by global disturbances, such as mergers or close encounters, 
which can remove the angular-momentum barrier, 
funnel gas inward to form a gas-rich, compact, supernova-free nucleus, 
and sustain the super-Eddington BH growth and nuclear star formation 
over a short timescale of $\lesssim {\rm Myr}$.
The origin of such global disturbances is tied to excursions to a fast
phase in the assembly histories of host halos, suggesting that the LRD population may be a natural
consequence of hierarchical structure formation in the $\Lambda$CDM cosmology.

Our model is among the first to self-consistently include the formation of seeds 
and the BH-galaxy-halo co-evolution within a cosmological context, allowing the emergence 
of the LRD population as a natural outcome of the $\Lambda$CDM paradigm, rather 
than a result of fine-tuning. 
In fact, none of the model prescriptions or parameters
was designed or tuned according to the observations of the LRDs.
The strength of the halo excursion required to trigger a nuclear burst
is set to the same value that separates the two phases of 
$\Lambda$CDM halos (see Eq.~C3 of 
\citetalias{chenTwophaseModelGalaxy2025a} and \S2 of 
\citetalias{moTwophaseModelGalaxy2024}); 
the efficiency of star formation in nuclear star bursts is set by the 
free-fall timescale and the BH accretion there is modeled on-the-fly 
using a turbulence-modified Bondi accretion plus a turbulent and 
magnetized accretion disk (see Appendix C of 
\citetalias{chenTwophaseModelGalaxy2025a});
the mass loading of AGN feedback in nuclear bursts is determined by 
the Eddington-level luminosity and a typical outflow velocity of 
$10^3\kms$ (see Appendix C5 of \citetalias{chenTwophaseModelGalaxy2025a}). 

The LRD population predicted by our model reproduces a broad range of 
observations, and our model can make predictions for 
many more properties of the LRD population themselves, as well as their 
progenitors and descendants
(\S\ref{sec:predictions}; Figs.~\ref{fig:n_vs_z} and \ref{fig:prop_dist}).
This allows the prescriptions implemented in the model to be tested by 
observations targeting different evolution stages of LRDs
and the differences from other models to be explicitly examined.
We discuss some of these aspects below.

Although the BH$^{\star}$ scenario has previously been suggested 
and accepted as a promising interpretation of LRDs, our model provides 
a physical framework to explain how the conditions required for 
BH$^{\star}$ formation are created in the galaxy formation context.
Our model differs from other models that interpret LRDs as (DC)BHs formed 
in the first collapses of the halo gas \citep{baggenConnectingDotsUVbright2026}.
For example, the LRDs at $z = 5$ predicted by our model are seeded at $z \gtrsim 20$ 
and have since gone through $\sim 10$ episodes of nuclear bursts 
(see Fig.~9 of \citetalias{chenTwophaseModelGalaxy2025a})
interspersed with phases of continuous growth.
Our model differs from that suggested by 
\citet{inayoshiSpectralUniformityLittle2026} in the following aspects:
the shining phase of an LRD in our model is a violent and bursty event,
episodically triggered by global disturbance, 
asynchronous with the evolution of their host galaxies, 
and lasting for $\lesssim 1\Myr$, instead of a secular process over 
$\gtrsim 10\Myr$ in the nuclear region extrapolated from 
the host galactic disk; the observed LRDs in our model are 
objects with significant post-seeding growth, instead of nascent BHs embedded 
in nascent star-forming galaxies; the depletion of gas in nuclear regions is 
mostly driven by the feedback from the accreting BHs, rather than stellar feedback.

A unique consequence of episodic bursts in our model is that 
the LRDs currently observed are just the tip of the iceberg shaped by 
bursty-mode growth (see Fig.~\ref{fig:selection}a).
Observations with improved sensitivity are thus expected to reveal 
many more BHs below the waterline of the current detection limit.
Another consequence is that the companion NSC of an LRD should harbor a 
significant fraction of ancient stars formed 
through previous nuclear bursts since $z \gtrsim 20$ and, at the same time,
possess a large fraction of young stars with age $\lesssim 1\Myr$ formed during 
the ongoing burst. This prediction may be tested by observations that can resolve 
individual stellar populations through SED or spatial decompositions.
The ``about-to-be-built'' extended stellar components in the host galaxies 
of LRDs also suggest that some observed LRDs may already have built 
up extended stellar components that can be revealed by stacking 
analysis and by deep imaging observations \citep[e.g.][]{zhangUnveilingExtendedComponents2025,cloonanPANORAMICUVopticalMorphologies2026}.

The complete coverage of the evolution history of the LRD population
by our model allows us to seek observational evidence for both the seeding and 
growing processes through archaeology of their descendants at lower $z$.
For example, the mass gap of the BH-seed mass function produced by PISNe may
lead to a bimodal distribution of galaxies in their post-seeding 
star formation, provided that the effects of 
BH feedback depend on BH mass. Observations targeting galaxies at $z \gtrsim 10$
\citep[e.g.][]{maiolinoSmallVigorousBlack2024,tacchellaJADESImagingGNz112023,jiJADESSmallBlue2025,bogdanEvidenceHeavyseedOrigin2023,naiduCosmicMiracleRemarkably2026,carnianiSpectroscopicConfirmationTwo2024,curtis-lakeSpectroscopicConfirmationFour2023}, before the evolution of BHs and their 
host galaxies converges due to self-regulation 
(see \S3.3.2 and Fig.~11 of \citetalias{chenTwophaseModelGalaxy2025a}), 
may be able to identify such a bimodality.
Another implication of the model is that LRDs seeded after PISNe should be 
associated with stars enriched by the ejecta of PISNe. 
The peculiar abundance patterns of PISN enrichment, as seen in a small 
set of stars discovered in the local Universe \citep[e.g.][]{xingMetalpoorStarAbundances2023,skuladottirPairinstabilitySupernovaOrigin2024}, may 
thus be used to link the descendants of LRDs to their seeding pathways.

As shown by the blue histogram in Fig.~\ref{fig:prop_dist}g--i, 
our model predicts that a significant fraction of the LRD population 
gained only modest amounts in mass and the stellar mass of 
the descendants matches that of dwarf galaxies in the local Universe.
These descendants are predicted to harbor BHs with $M_{\rm BH}$
similar to those of LRDs and to have a significant fraction of $M_\star$ 
formed during the early bursty growth. In particular, the absence of 
large amounts of growth suggests that such a descendant should contain 
a compact component formed through early nuclear bursts, reminiscent of 
an ultra-compact dwarf (UCD) or the central part of a nucleated dwarf 
\citep[e.g.][]{wangEvolutionaryContinuumNucleated2023}. 
We will present a detailed analysis of the connection between LRDs and present-day 
compact dwarf galaxies in a forthcoming paper. 



\begin{acknowledgments}
This work is supported by 
Double First-Class Discipline Construction-Astronomy Discipline at Nanjing University,
the National Natural Science Foundation of China (Grant no. 12503014),
and the Fundamental Research Funds for the Central Universities (Grant no. KG202502).
The authors thank Tao Wang, Enci Wang and Zhaozhou Li for their discussions
and insights, and thank Yi Mao and Chen Chen for their technical support.
The authors would like to express their gratitude to the Tsinghua Astrophysics 
High-Performance Computing platform at Tsinghua University for 
providing the necessary computational and data storage resources that have 
significantly contributed to the research results presented in this paper.
The authors thank the anonymous referee for their valuable comments.
\end{acknowledgments}

\begin{contribution}

The authors contributed equally to this work.


\end{contribution}

%

\software{\textsc{Hipp} \citep{chenHIPPHIghPerformancePackage2023},
\textsc{Astropy} \citep{robitailleAstropyCommunityPython2013,
astropycollaborationAstropyProjectBuilding2018,
astropycollaborationAstropyProjectSustaining2022},
\textsc{SatGen}\citep{jiangSatGenSemianalyticalSatellite2021},
\textsc{WebPlotDigitizer}}.

\bigskip
{\large \it Data availability: } 
Codes implementing our model, catalogs produced by the model, and data and scripts 
to reproduce all figures will be available at the repository 
\textsc{TwoPhaseGalaxyModel} (\url{https://github.com/ChenYangyao/two-phase-galaxy-model}). 
The N-body data have been made public by the \textsc{IllustrisTNG} 
project (\url{https://www.tng-project.org/}).


\appendix

\section{Bolometric luminosity functions}
\label{app:sec:lf}

\begin{figure*} \centering
    \includegraphics[width=0.8\textwidth]{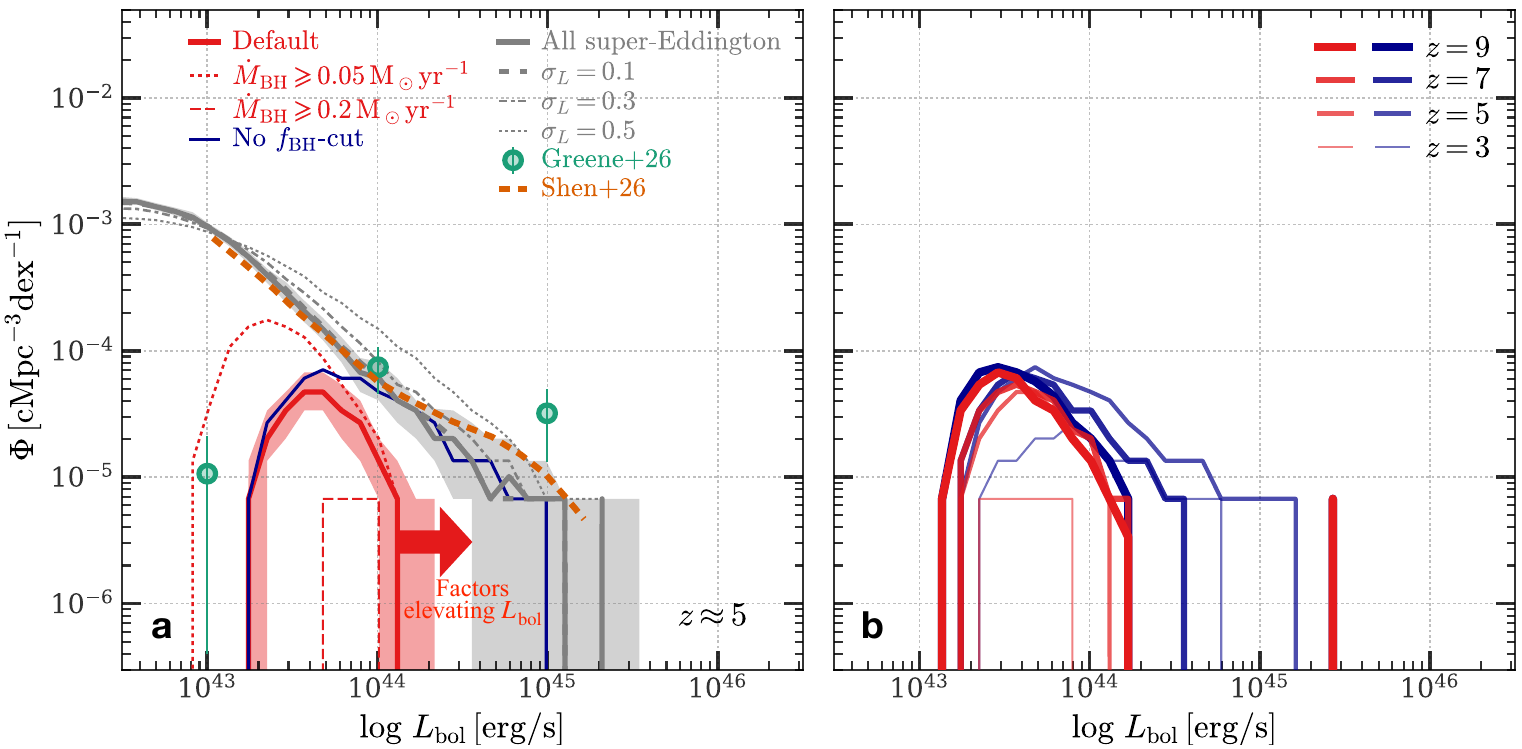}
    \caption{{\figem Bolometric luminosity function ($\Phi$) of LRDs.}
    {\figem a}, $\Phi$ of LRDs predicted by our model at $z=5$ using the 
    default selection ({\figem red solid}) and using a number of
    alternative selections adapted from the default one (see \S\ref{ssec:sample}):
    varying the thresholds of $\dot{M}_{\rm BH}$ 
    ({\figem red dashed, dotted and dash-dotted});
    removing the $f_{\rm BH}$ threshold ({\figem blue}; 
    i.e. by additionally including the blue points in Fig.~\ref{fig:selection});
    retaining only the super-Eddington requirement ({\figem grey solid};
    labeled ``All super-Eddington'').
    We also show the predictions for all super-Eddington BHs
    by adding log-normal scatter with different widths to their 
    $L_{\rm bol}$ ({\figem grey dashed, dash-dotted and dotted}). 
    In each case, the curve represents the median, and shading (for the default and all super-Eddington cases) 
    represents the $16^{\rm th}$--$84^{\rm th}$ percentile range among 
    100 random samples for the durations of nuclear bursts.
    For comparison, we show the results from the literature: 
    the observed LRDs at $z = 4$--$6$ with a new bolometric correction 
    (\citealt{greeneWhatYouSee2026}; {\figem green}); 
    the LRDs modeled by an empirical mapping from the 
    BHs in the \textsc{Lumina} simulation
    (\citealt{shenLuminaProjectDemographics2026}; {\figem orange}).
    The prediction for $\Phi$ is subject to uncertain factors
    that may elevate $L_{\rm bol}$, as indicated by a 
    {\figem red arrow} (see Appendix~\ref{app:sec:lf} for discussion).
    {\figem b}, $\Phi$ of LRDs predicted by our model 
    at different redshifts, using the default selection ({\figem red}) 
    and the selection without the $f_{\rm BH}$ threshold ({\figem blue}).
    }
    \label{fig:lf}
\end{figure*}

As discussed in \S\ref{ssec:sample}, the bolometric luminosity of a BH in 
super-Eddington accretion is expected to be limited around the Eddington 
luminosity, $L_{\rm Edd}$. We thus use this assumption to predict their 
bolometric luminosities as $L_{\rm bol} = L_{\rm Edd}$.
Fig.~\ref{fig:lf}a shows the predicted bolometric luminosity function,
$\Phi$, defined as the number of LRDs per 
unit logarithmic bolometric luminosity per unit 
comoving volume, obtained from different selections at $z = 5$.
With the default selection, the predicted $\Phi(L_{\rm bol})$ reaches a peak
of $\sim 5 \times 10^{-5}\,{\rm cMpc^{-3}dex^{-1}}$ 
at $L_{\rm bol} \sim 5 \times 10^{44}\,{\rm erg/s}$.
Tests with alternative selections show how the predicted $\Phi(L_{\rm bol})$ 
depends on the selection criteria:
varying the threshold of $\dot{M}_{\rm BH}$ while keeping the other
selection criteria unchanged changes the number of LRDs at the faint 
end, expected from the fact that $L_{\rm Edd}$ is proportional to $M_{\rm BH}$
and $\dot{M}_{\rm BH}$ tightly correlates with $M_{\rm BH}$ 
for our super-Eddington branch (see Fig.~\ref{fig:selection}a);
removing the $f_{\rm BH}$ threshold broadens the luminosity function 
towards the bright end, as LRDs are the population most readily able to 
build a significant stellar component in an extended configuration
through the continuous mode of star formation,
and those with higher $M_{\rm BH}$ are more likely to achieve this 
(see the transition of $f_{\rm BH}$ at $M_{\rm BH} \sim 10^6$--$10^7\Msun$ 
in Fig.~\ref{fig:selection}b and the discussion in 
\S\ref{ssec:prop_dist});
including all BHs in super-Eddington accretion extends the luminosity 
function to both the faint and bright ends.

For comparison, in Fig.~\ref{fig:lf}a, we show the bolometric luminosity functions 
obtained in the literature. Before drawing a conclusion, we note that our model, 
including both the seeding and growing procedures, was only calibrated to match  
the $M_{\rm BH}$-$M_\star$ and $M_\star$-$M_{\rm halo}$ relations
at $z \approx 0$ (see Appendix D of \citetalias{moTwophaseModelGalaxy2024})
without using any observational information at high $z$. Our predictions for 
the LRD population thus serve as a blind test.

The bolometric luminosity function of observed LRDs at $z = 4$--$6$ recently 
obtained by \citet[green in Fig.~\ref{fig:lf}a]{greeneWhatYouSee2026} 
using an empirically derived bolometric correction shows a peak of 
$\sim 10^{-4}\,{\rm cMpc^{-3}dex^{-1}}$ 
at $L_{\rm bol} \sim 10^{44}\,{\rm erg/s}$, close to our prediction
with the default selection. 
Their $\Phi$ at the bright end, however, is more compatible with our prediction 
when the $f_{\rm BH}$ threshold is removed. 
This indicates that the observed LRDs may contain a population whose 
extended stellar components are already built up -- a debatable conclusion that 
may be answered through stacking analyses in observations
\citep[e.g.][]{zhangUnveilingExtendedComponents2025,
cloonanPANORAMICUVopticalMorphologies2026}. 

Given the uncertainties in both the observations and the model, there are 
other factors that may elevate the predicted $\Phi(L_{\rm bol})$ and explain the 
difference between the predictions and those obtained by \citet{greeneWhatYouSee2026}. First, the bolometric correction used by \citet{greeneWhatYouSee2026} is empirically derived from a small number of LRDs, 
which does not necessarily account for the variations in the SEDs of the 
entire population of LRDs.
Second, factors such as cosmic variance, intrinsic variations in $L_{\rm bol}$ 
of BHs, and observational uncertainties in sample selection and 
photometric measurements, can introduce a certain level of scatter 
into the observed $L_{\rm bol}$ and broaden the resulting $\Phi(L_{\rm bol})$
towards the bright end through Eddington bias \citep[e.g.][]{eddingtonFormulaCorrectingStatistics1913,chenMassiveDarkMatter2023,shenImpactUVVariability2023}.
This effect can be seen from the grey curves in Fig.~\ref{fig:lf}a, 
for which different levels of log-normal scatter are added 
to the predicted $L_{\rm bol}$.
Third, the luminosity of a BH in super-Eddington accretion 
may be over-simplified by the assumption of $L_{\rm bol} = L_{\rm Edd}$
in our model. In fact, there are theoretical studies suggesting
a logarithmic increase of $L_{\rm bol}$ with $\dot{M}_{\rm BH}$
in the super-Eddington regime. At $\dot{M}_{\rm BH}/\dot{M}_{\rm Edd} \sim 10$ 
predicted by our model for the super-Eddington branch 
(Fig.~\ref{fig:selection}a), this can elevate 
$L_{\rm bol}$ by a factor of a few above $L_{\rm Edd}$
(see e.g. Eq.~25 of \citealt{lipunovaSupercriticalDiskAccretion1999} and Fig.~5 of \citealt{kubotaModellingSpectralEnergy2019}).
Finally, the continuous-mode growth (see \S\ref{ssec:model}), 
which accounts for the capturing of sub-clouds formed in the turbulence,
is modeled in a smoothed manner that takes into account the 
time-averaged capture rate. While the variations in $\dot{M}_{\rm BH}$ are 
expected to be smaller than those in the bursty mode, capturing an individual sub-cloud 
may introduce a rise and fall in $\dot{M}_{\rm BH}$ and, in extreme cases, 
lead to conditions capable of powering an LRD.
Modeling such a process requires tracking individual sub-clouds and their 
interactions with the BH, which can be explored by extending the framework of
Monte-Carlo sampling of sub-clouds developed by 
\citet{chenTwophaseModelGalaxy2025}.

The bolometric luminosity function of LRDs predicted by
\citet[orange in Fig.~\ref{fig:lf}a]{shenLuminaProjectDemographics2026} at $z \approx 5$ 
using an empirical mapping from the BH population in the \textsc{Lumina} 
simulation is broadly consistent with our prediction including all 
super-Eddington BHs,
indicating that the two populations are physically equivalent.
In fact, \citet{shenLuminaProjectDemographics2026} selected LRDs as the SMBHs 
whose masses are close to the seeded mass 
($M_{\rm BH} \lesssim 10M_{\rm BH, seed} \sim 10^7\Msun$), with a duty-cycle 
parameter of $\sim 0.3$ to account for their shining episodes.
In our model, the seeding procedure places BH seeds (mostly IMBHs) with $M_{\rm BH}$ 
much smaller than the observed $M_{\rm BH}$ of 
LRDs, and the post-seeding growth via episodic nuclear bursts drives these 
seeds to the regime of SMBHs (see \S3.2.3 and \S3.3 of \citetalias{chenTwophaseModelGalaxy2025a}).
Therefore, our model provides a physical explanation for the SMBH seeds 
placed in hydrodynamical simulations and the duty-cycles of their post-seeding 
growth, and interprets the LRDs as the last few shining episodes 
after which nuclear bursts cease.

Finally, we show the evolution of the bolometric luminosity function of LRDs 
predicted by our model in Fig.~\ref{fig:lf}b. With the default selection, 
the predicted $\Phi(L_{\rm bol})$ shows similar shape and amplitude at $z \gtrsim 5$,
but then drops sharply in amplitude at lower $z$, consistent with the 
evolution of the number density of LRDs shown in Fig.~\ref{fig:n_vs_z}.
Removing the $f_{\rm BH}$ threshold leads to a higher $\Phi(L_{\rm bol})$ at the bright end,
and more so at lower $z$, indicating that more significant
extended stellar components have been built up in the host galaxies of 
LRDs at lower $z$. This also implies an evolution of the $M_{\rm BH}$-$M_\star$ 
relation of SMBHs predicted by our model, 
as shown by \citetalias{chenTwophaseModelGalaxy2025a} (see their Fig. 12).
At $z = 2$, the entire simulation volume contains only $\sim 2$ LRDs by the default selection,
resulting in large Poisson noise in the predicted $\Phi(L_{\rm bol})$. We thus do not 
show the predicted $\Phi(L_{\rm bol})$ at $z \leqslant 2$, and leave the study of LRDs 
at lower $z$ to future work with larger simulation volumes.


\bibliography{ref}{}

\begin{thebibliography}{}
\expandafter\ifx\csname natexlab\endcsname\relax\def\natexlab#1{#1}\fi
\providecommand{\url}[1]{\href{#1}{#1}}
\providecommand{\dodoi}[1]{doi:~\href{http://doi.org/#1}{\nolinkurl{#1}}}
\providecommand{\doeprint}[1]{\href{http://ascl.net/#1}{\nolinkurl{http://ascl.net/#1}}}
\providecommand{\doarXiv}[1]{\href{https://arxiv.org/abs/#1}{\nolinkurl{https://arxiv.org/abs/#1}}}

\bibitem[{H.~B. Akins {et~al.}(2025)Akins, Casey, Lambrides, Allen, Andika,
  Brinch, Champagne, Cooper, Ding, Drakos, Faisst, Finkelstein, Franco,
  Fujimoto, Gentile, Gillman, Gozaliasl, Harish, Hayward, Hirschmann, Ilbert,
  Kartaltepe, Kocevski, Koekemoer, Kokorev, Liu, Long, McCracken, McKinney,
  Onoue, Paquereau, Renzini, Rhodes, Robertson, Shuntov, Silverman, Tanaka,
  Toft, Trakhtenbrot, Valentino, \&
  Zavala}]{akinsCOSMOSWebOverabundancePhysical2025}
Akins, H.~B., Casey, C.~M., Lambrides, E., {et~al.} 2025,
  \bibinfo{title}{{{COSMOS-Web}}: {{The Overabundance}} and {{Physical Nature}}
  of "{{Little Red Dots}}"---{{Implications}} for {{Early Galaxy}} and {{SMBH
  Assembly}},} The Astrophysical Journal, 991, 37,
  \dodoi{10.3847/1538-4357/ade984}

\bibitem[{Y. Asada {et~al.}(2024)Asada, Sawicki, Abraham, Brada{\v c}, Brammer,
  Desprez, {Estrada-Carpenter}, Iyer, Martis, Matharu, Mowla, Muzzin, Noirot,
  Sarrouh, Strait, Willott, \& Harshan}]{asadaBurstyStarFormation2024}
Asada, Y., Sawicki, M., Abraham, R., {et~al.} 2024, \bibinfo{title}{Bursty Star
  Formation and Galaxy-Galaxy Interactions in Low-Mass Galaxies 1 {{Gyr}} after
  the {{Big Bang}},} Monthly Notices of the Royal Astronomical Society, 527,
  11372, \dodoi{10.1093/mnras/stad3902}

\bibitem[{ {Astropy Collaboration} {et~al.}(2018){Astropy Collaboration},
  {Price-Whelan}, Sip{\H o}cz, G{\"u}nther, Lim, Crawford, Conseil, Shupe,
  Craig, Dencheva, Ginsburg, VanderPlas, Bradley, {P{\'e}rez-Su{\'a}rez}, {de
  Val-Borro}, Aldcroft, Cruz, Robitaille, Tollerud, Ardelean, Babej, Bach,
  Bachetti, Bakanov, Bamford, Barentsen, Barmby, Baumbach, Berry, Biscani,
  Boquien, Bostroem, Bouma, Brammer, Bray, Breytenbach, Buddelmeijer, Burke,
  Calderone, Cano~Rodr{\'i}guez, Cara, Cardoso, Cheedella, Copin, Corrales,
  Crichton, D'Avella, Deil, Depagne, Dietrich, Donath, Droettboom, Earl, Erben,
  Fabbro, Ferreira, Finethy, Fox, Garrison, Gibbons, Goldstein, Gommers, Greco,
  Greenfield, Groener, Grollier, Hagen, Hirst, Homeier, Horton, Hosseinzadeh,
  Hu, Hunkeler, Ivezi{\'c}, Jain, Jenness, Kanarek, Kendrew, Kern, Kerzendorf,
  Khvalko, King, Kirkby, Kulkarni, Kumar, Lee, Lenz, Littlefair, Ma, Macleod,
  Mastropietro, McCully, Montagnac, Morris, Mueller, Mumford, Muna, Murphy,
  Nelson, Nguyen, Ninan, N{\"o}the, Ogaz, Oh, Parejko, Parley, Pascual, Patil,
  Patil, Plunkett, Prochaska, Rastogi, Reddy~Janga, Sabater, Sakurikar,
  Seifert, Sherbert, {Sherwood-Taylor}, Shih, Sick, Silbiger, Singanamalla,
  Singer, Sladen, Sooley, Sornarajah, Streicher, Teuben, Thomas, Tremblay,
  Turner, Terr{\'o}n, {van Kerkwijk}, {de la Vega}, Watkins, Weaver, Whitmore,
  Woillez, Zabalza, \& {Astropy
  Contributors}}]{astropycollaborationAstropyProjectBuilding2018}
{Astropy Collaboration}, {Price-Whelan}, A.~M., Sip{\H o}cz, B.~M., {et~al.}
  2018, \bibinfo{title}{The {{Astropy Project}}: {{Building}} an {{Open-science
  Project}} and {{Status}} of the v2.0 {{Core Package}},} The Astronomical
  Journal, 156, 123, \dodoi{10.3847/1538-3881/aabc4f}

\bibitem[{ {Astropy Collaboration} {et~al.}(2022){Astropy Collaboration},
  {Price-Whelan}, Lim, Earl, Starkman, Bradley, Shupe, Patil, Corrales,
  Brasseur, N{\"o}the, Donath, Tollerud, Morris, Ginsburg, Vaher, Weaver,
  Tocknell, Jamieson, {van Kerkwijk}, Robitaille, Merry, Bachetti, G{\"u}nther,
  Aldcroft, {Alvarado-Montes}, Archibald, B{\'o}di, Bapat, Barentsen,
  Baz{\'a}n, Biswas, Boquien, Burke, Cara, Cara, Conroy, Conseil, Craig, Cross,
  Cruz, D'Eugenio, Dencheva, Devillepoix, Dietrich, Eigenbrot, Erben, Ferreira,
  {Foreman-Mackey}, Fox, Freij, Garg, Geda, Glattly, Gondhalekar, Gordon,
  Grant, Greenfield, Groener, Guest, Gurovich, Handberg, Hart,
  {Hatfield-Dodds}, Homeier, Hosseinzadeh, Jenness, Jones, Joseph, Kalmbach,
  Karamehmetoglu, Ka{\l}uszy{\'n}ski, Kelley, Kern, Kerzendorf, Koch, Kulumani,
  Lee, Ly, Ma, MacBride, Maljaars, Muna, Murphy, Norman, O'Steen, Oman,
  Pacifici, Pascual, {Pascual-Granado}, Patil, Perren, Pickering, Rastogi,
  Roulston, Ryan, Rykoff, Sabater, Sakurikar, Salgado, Sanghi, Saunders,
  Savchenko, Schwardt, {Seifert-Eckert}, Shih, Jain, Shukla, Sick, Simpson,
  Singanamalla, Singer, Singhal, Sinha, Sip{\H o}cz, Spitler, Stansby,
  Streicher, {\v S}umak, Swinbank, Taranu, Tewary, Tremblay, {de Val-Borro},
  Van~Kooten, Vasovi{\'c}, Verma, {de Miranda Cardoso}, Williams, Wilson,
  Winkel, {Wood-Vasey}, Xue, Yoachim, Zhang, Zonca, \& {Astropy Project
  Contributors}}]{astropycollaborationAstropyProjectSustaining2022}
{Astropy Collaboration}, {Price-Whelan}, A.~M., Lim, P.~L., {et~al.} 2022,
  \bibinfo{title}{The {{Astropy Project}}: {{Sustaining}} and {{Growing}} a
  {{Community-oriented Open-source Project}} and the {{Latest Major Release}}
  (v5.0) of the {{Core Package}},} The Astrophysical Journal, 935, 167,
  \dodoi{10.3847/1538-4357/ac7c74}

\bibitem[{J.~F.~W. Baggen {et~al.}(2026)Baggen, Scoggins, {van Dokkum}, Haiman,
  Torralba, \& Matthee}]{baggenConnectingDotsUVbright2026}
Baggen, J. F.~W., Scoggins, M.~T., {van Dokkum}, P., {et~al.} 2026,
  \bibinfo{title}{Connecting the {{Dots}}: {{UV-bright Companions}} of {{Little
  Red Dots}} as {{Lyman}}-{{Werner Sources Enabling Direct-collapse Black Hole
  Formation}},} The Astrophysical Journal, 1002, L4,
  \dodoi{10.3847/2041-8213/ae58a5}

\bibitem[{M.~C. Begelman \& J. Dexter(2026)Begelman \&
  Dexter}]{begelmanLittleRedDots2026}
Begelman, M.~C., \& Dexter, J. 2026, \bibinfo{title}{Little {{Red Dots}} as
  {{Late-stage Quasi-stars}},} The Astrophysical Journal, 996, 48,
  \dodoi{10.3847/1538-4357/ae274a}

\bibitem[{{\'A}. Bogd{\'a}n {et~al.}(2023)Bogd{\'a}n, Goulding, Natarajan,
  Kov{\'a}cs, Tremblay, Chadayammuri, Volonteri, Kraft, Forman, Jones,
  Churazov, \& Zhuravleva}]{bogdanEvidenceHeavyseedOrigin2023}
Bogd{\'a}n, {\'A}., Goulding, A.~D., Natarajan, P., {et~al.} 2023,
  \bibinfo{title}{Evidence for Heavy-Seed Origin of Early Supermassive Black
  Holes from a z {$\approx$} 10 {{X-ray}} Quasar,} Nature Astronomy, 8, 126,
  \dodoi{10.1038/s41550-023-02111-9}

\bibitem[{S. Carniani {et~al.}(2024)Carniani, Hainline, D'Eugenio, Eisenstein,
  Jakobsen, Witstok, Johnson, Chevallard, Maiolino, Helton, Willott, Robertson,
  Alberts, Arribas, Baker, Bhatawdekar, Boyett, Bunker, Cameron, Cargile,
  Charlot, Curti, {Curtis-Lake}, Egami, Giardino, Isaak, Ji, Jones, Kumari,
  Maseda, Parlanti, {P{\'e}rez-Gonz{\'a}lez}, Rawle, Rieke, Rieke, Del~Pino,
  Saxena, Scholtz, Smit, Sun, Tacchella, {\"U}bler, Venturi, Williams, \&
  Willmer}]{carnianiSpectroscopicConfirmationTwo2024}
Carniani, S., Hainline, K., D'Eugenio, F., {et~al.} 2024,
  \bibinfo{title}{Spectroscopic Confirmation of Two Luminous Galaxies at a
  Redshift of 14,} Nature, 633, 318, \dodoi{10.1038/s41586-024-07860-9}

\bibitem[{M. {Carranza-Escudero} {et~al.}(2025){Carranza-Escudero}, Conselice,
  Adams, Harvey, Austin, Behroozi, Ferreira, Ormerod, Duan, Trussler, Li,
  Westcott, Windhorst, Coe, Cohen, Cheng, Driver, Frye, Furtak, Grogin, Hathi,
  Jansen, Koekemoer, Marshall, O'Brien, Pirzkal, Polletta, Robotham, Rutkowski,
  Summers, Wilkins, Willmer, Yan, \&
  Zitrin}]{carranza-escuderoLonelyLittleRed2025}
{Carranza-Escudero}, M., Conselice, C.~J., Adams, N., {et~al.} 2025,
  \bibinfo{title}{Lonely {{Little Red Dots}}: {{Challenges}} to the {{Active
  Galactic Nucleus Nature}} of {{Little Red Dots}} through {{Their Clustering}}
  and {{Spectral Energy Distributions}},} The Astrophysical Journal, 989, L50,
  \dodoi{10.3847/2041-8213/adf73d}

\bibitem[{G. Chabrier(2003)Chabrier}]{chabrierGalacticStellarSubstellar2003}
Chabrier, G. 2003, \bibinfo{title}{Galactic {{Stellar}} and {{Substellar
  Initial Mass Function1}},} Publications of the Astronomical Society of the
  Pacific, 115, 763, \dodoi{10.1086/376392}

\bibitem[{C.-H. Chen {et~al.}(2025)Chen, Ho, Li, \&
  Zhuang}]{chenHostGalaxyIf2025}
Chen, C.-H., Ho, L.~C., Li, R., \& Zhuang, M.-Y. 2025, \bibinfo{title}{The
  {{Host Galaxy}} ({{If Any}}) of the {{Little Red Dots}},} The Astrophysical
  Journal, 983, 60, \dodoi{10.3847/1538-4357/ada93a}

\bibitem[{Y. Chen {et~al.}(2024)Chen, Mo, \&
  Wang}]{chenTwophaseModelGalaxy2024}
Chen, Y., Mo, H., \& Wang, H. 2024, \bibinfo{title}{A Two-Phase Model of Galaxy
  Formation - {{II}}. {{The}} Size-Mass Relation of Dynamically Hot Galaxies,}
  Monthly Notices of the Royal Astronomical Society, 532, 4340,
  \dodoi{10.1093/mnras/stae1757}

\bibitem[{Y. Chen {et~al.}(2025{\natexlab{a}})Chen, Mo, \&
  Wang}]{chenTwophaseModelGalaxy2025a}
Chen, Y., Mo, H., \& Wang, H. 2025{\natexlab{a}}, A Two-Phase Model of Galaxy
  Formation: {{IV}}. {{Seeding}} and Growing Supermassive Black Holes in Dark
  Matter Halos, arXiv, \dodoi{10.48550/arXiv.2509.03283}

\bibitem[{Y. Chen {et~al.}(2025{\natexlab{b}})Chen, Mo, \&
  Wang}]{chenTwophaseModelGalaxy2025}
Chen, Y., Mo, H., \& Wang, H. 2025{\natexlab{b}}, \bibinfo{title}{A Two-Phase
  Model of Galaxy Formation: {{III}}. {{The}} Formation of Globular Clusters,}
  Monthly Notices of the Royal Astronomical Society, 540, 1235,
  \dodoi{10.1093/mnras/staf791}

\bibitem[{Y. Chen {et~al.}(2019)Chen, Mo, Li, Wang, Yang, Zhou, \&
  Zhang}]{chenELUCIDVICosmic2019}
Chen, Y., Mo, H.~J., Li, C., {et~al.} 2019, \bibinfo{title}{{{ELUCID}}. {{VI}}.
  {{Cosmic Variance}} of the {{Galaxy Distribution}} in the {{Local
  Universe}},} The Astrophysical Journal, 872, 180,
  \dodoi{10.3847/1538-4357/ab0208}

\bibitem[{Y. Chen {et~al.}(2023{\natexlab{a}})Chen, Mo, Li, Wang, Wang, \&
  Yang}]{chenConditionalAbundanceMatching2023}
Chen, Y., Mo, H.~J., Li, C., {et~al.} 2023{\natexlab{a}}, \bibinfo{title}{A
  Conditional Abundance Matching Method of Extending Simulated Halo Merger
  Trees to Resolve Low-Mass Progenitors and Subhalos,} Monthly Notices of the
  Royal Astronomical Society, 525, 1254, \dodoi{10.1093/mnras/stad2336}

\bibitem[{Y. Chen {et~al.}(2023{\natexlab{b}})Chen, Mo, \&
  Wang}]{chenMassiveDarkMatter2023}
Chen, Y., Mo, H.~J., \& Wang, K. 2023{\natexlab{b}}, \bibinfo{title}{Massive
  Dark Matter Haloes at High Redshift: Implications for Observations in the
  {{JWST}} Era,} Monthly Notices of the Royal Astronomical Society, 526, 2542,
  \dodoi{10.1093/mnras/stad2866}

\bibitem[{Y. Chen \& K. Wang(2023)Chen \&
  Wang}]{chenHIPPHIghPerformancePackage2023}
Chen, Y., \& Wang, K. 2023, \bibinfo{title}{{{HIPP}}: {{HIgh-Performance
  Package}} for Scientific Computation,} Astrophysics Source Code Library,
  ascl:2301.030.
\newblock \url{https://ui.adsabs.harvard.edu/abs/2023ascl.soft01030C}

\bibitem[{A.~P. Cloonan {et~al.}(2026)Cloonan, Whitaker, Manning, Williams,
  Greene, Oesch, Weibel, Brammer, {de Graaff}, Hviding, Dayal, Jespersen, Ji,
  Labbe, Xiao, \& Zhang}]{cloonanPANORAMICUVopticalMorphologies2026}
Cloonan, A.~P., Whitaker, K.~E., Manning, S.~M., {et~al.} 2026,
  \bibinfo{title}{A {{PANORAMIC}} of {{UV-optical}} Morphologies of "{{Little
  Red Dots}}": {{Two}} Groups of {{LRDs}} Distinguished by {{UV}} Half-Light
  Radius,} arXiv e-prints, arXiv:2603.24700, \dodoi{10.48550/arXiv.2603.24700}

\bibitem[{C. Conroy \& J.~E. Gunn(2010)Conroy \&
  Gunn}]{conroyPropagationUncertaintiesStellar2010}
Conroy, C., \& Gunn, J.~E. 2010, \bibinfo{title}{The {{Propagation}} of
  {{Uncertainties}} in {{Stellar Population Synthesis Modeling}}. {{III}}.
  {{Model Calibration}}, {{Comparison}}, and {{Evaluation}},} The Astrophysical
  Journal, 712, 833, \dodoi{10.1088/0004-637X/712/2/833}

\bibitem[{C. Conroy {et~al.}(2009)Conroy, Gunn, \&
  White}]{conroyPropagationUncertaintiesStellar2009}
Conroy, C., Gunn, J.~E., \& White, M. 2009, \bibinfo{title}{The {{Propagation}}
  of {{Uncertainties}} in {{Stellar Population Synthesis Modeling}}. {{I}}.
  {{The Relevance}} of {{Uncertain Aspects}} of {{Stellar Evolution}} and the
  {{Initial Mass Function}} to the {{Derived Physical Properties}} of
  {{Galaxies}},} The Astrophysical Journal, 699, 486,
  \dodoi{10.1088/0004-637X/699/1/486}

\bibitem[{R.~A. Crain {et~al.}(2015)Crain, Schaye, Bower, Furlong, Schaller,
  Theuns, Dalla~Vecchia, Frenk, McCarthy, Helly, Jenkins, {Rosas-Guevara},
  White, \& Trayford}]{crainEAGLESimulationsGalaxy2015}
Crain, R.~A., Schaye, J., Bower, R.~G., {et~al.} 2015, \bibinfo{title}{The
  {{EAGLE}} Simulations of Galaxy Formation: Calibration of Subgrid Physics and
  Model Variations,} Monthly Notices of the Royal Astronomical Society, 450,
  1937, \dodoi{10.1093/mnras/stv725}

\bibitem[{E. {Curtis-Lake} {et~al.}(2023){Curtis-Lake}, Carniani, Cameron,
  Charlot, Jakobsen, Maiolino, Bunker, Witstok, Smit, Chevallard, Willott,
  Ferruit, Arribas, Bonaventura, Curti, D'Eugenio, Franx, Giardino, Looser,
  L{\"u}tzgendorf, Maseda, Rawle, Rix, {Rodr{\'i}guez del Pino}, {\"U}bler,
  Sirianni, Dressler, Egami, Eisenstein, Endsley, Hainline, Hausen, Johnson,
  Rieke, Robertson, Shivaei, Stark, Tacchella, Williams, Willmer, Bhatawdekar,
  Bowler, Boyett, Chen, {de Graaff}, Helton, Hviding, Jones, Kumari, Lyu,
  Nelson, Perna, Sandles, Saxena, Suess, Sun, Topping, Wallace, \&
  Whitler}]{curtis-lakeSpectroscopicConfirmationFour2023}
{Curtis-Lake}, E., Carniani, S., Cameron, A., {et~al.} 2023,
  \bibinfo{title}{Spectroscopic Confirmation of Four Metal-Poor Galaxies at z =
  10.3-13.2,} Nature Astronomy, 7, 622, \dodoi{10.1038/s41550-023-01918-w}

\bibitem[{A. {de Graaff} {et~al.}(2025){de Graaff}, Rix, Naidu, Labb{\'e},
  Wang, Leja, Matthee, Katz, Greene, Hviding, Baggen, Bezanson, Boogaard,
  Brammer, Dayal, {van Dokkum}, Goulding, Hirschmann, Maseda, McConachie,
  Miller, Nelson, Oesch, Setton, Shivaei, Weibel, Whitaker, \&
  Williams}]{degraaffRemarkableRubyAbsorption2025}
{de Graaff}, A., Rix, H.-W., Naidu, R.~P., {et~al.} 2025, \bibinfo{title}{A
  Remarkable Ruby: {{Absorption}} in Dense Gas, Rather than Evolved Stars,
  Drives the Extreme {{Balmer}} Break of a Little Red Dot at z = 3.5,}
  Astronomy and Astrophysics, 701, A168, \dodoi{10.1051/0004-6361/202554681}

\bibitem[{F. Dong {et~al.}(2022)Dong, Zhao, Han, Li, Jing, \&
  Yang}]{dongUniversalSpecificMerger2022}
Dong, F., Zhao, D., Han, J., {et~al.} 2022, \bibinfo{title}{The {{Universal
  Specific Merger Rate}} of {{Dark Matter Halos}},} The Astrophysical Journal,
  929, 120, \dodoi{10.3847/1538-4357/ac5aaa}

\bibitem[{A.~S.
  Eddington(1913)Eddington}]{eddingtonFormulaCorrectingStatistics1913}
Eddington, A.~S. 1913, \bibinfo{title}{On a {{Formula}} for {{Correcting
  Statistics}} for the {{Effects}} of a Known {{Probable Error}} of
  {{Observation}},} Monthly Notices of the Royal Astronomical Society, 73, 359,
  \dodoi{10.1093/mnras/73.5.359}

\bibitem[{K. {El-Badry} {et~al.}(2016){El-Badry}, Wetzel, Geha, Hopkins,
  Kere{\v s}, Chan, \& {Faucher-Gigu{\`e}re}}]{el-badryBreathingFIREHow2016}
{El-Badry}, K., Wetzel, A., Geha, M., {et~al.} 2016, \bibinfo{title}{Breathing
  {{FIRE}}: {{How Stellar Feedback Drives Radial Migration}}, {{Rapid Size
  Fluctuations}}, and {{Population Gradients}} in {{Low-mass Galaxies}},} The
  Astrophysical Journal, 820, 131, \dodoi{10.3847/0004-637X/820/2/131}

\bibitem[{J.~E. Greene {et~al.}(2024)Greene, Labbe, Goulding, Furtak,
  Chemerynska, Kokorev, Dayal, Volonteri, Williams, Wang, Setton, Burgasser,
  Bezanson, Atek, Brammer, Cutler, Feldmann, Fujimoto, Glazebrook, {de Graaff},
  Khullar, Leja, Marchesini, Maseda, Matthee, Miller, Naidu, Nanayakkara,
  Oesch, Pan, Papovich, Price, {van Dokkum}, Weaver, Whitaker, \&
  Zitrin}]{greeneUNCOVERSpectroscopyConfirms2024}
Greene, J.~E., Labbe, I., Goulding, A.~D., {et~al.} 2024,
  \bibinfo{title}{{{UNCOVER Spectroscopy Confirms}} the {{Surprising Ubiquity}}
  of {{Active Galactic Nuclei}} in {{Red Sources}} at z {$>$} 5,} The
  Astrophysical Journal, 964, 39, \dodoi{10.3847/1538-4357/ad1e5f}

\bibitem[{J.~E. Greene {et~al.}(2026)Greene, Setton, Furtak, Naidu, Volonteri,
  Dayal, Labbe, {van Dokkum}, Bezanson, Brammer, Cutler, Glazebrook, {de
  Graaff}, Hirschmann, Hviding, Kokorev, Leja, Liu, Ma, Matthee, Nanayakkara,
  Oesch, Pan, Price, Spilker, Wang, Weaver, Whitaker, Williams, \&
  Zitrin}]{greeneWhatYouSee2026}
Greene, J.~E., Setton, D.~J., Furtak, L.~J., {et~al.} 2026,
  \bibinfo{title}{What You See Is What You Get: {{Empirically}} Measured
  Bolometric Luminosities of Little Red Dots,} The Astrophysical Journal, 996,
  129, \dodoi{10.3847/1538-4357/ae1836}

\bibitem[{T.~H. Greif {et~al.}(2011)Greif, White, Klessen, \&
  Springel}]{greifDelayPopulationIII2011}
Greif, T.~H., White, S. D.~M., Klessen, R.~S., \& Springel, V. 2011,
  \bibinfo{title}{The {{Delay}} of {{Population III Star Formation}} by
  {{Supersonic Streaming Velocities}},} The Astrophysical Journal, 736, 147,
  \dodoi{10.1088/0004-637X/736/2/147}

\bibitem[{P.~F. Hopkins {et~al.}(2023)Hopkins, Gurvich, Shen, Hafen,
  Grudi{\'c}, {Kurinchi-Vendhan}, Hayward, Jiang, Orr, Wetzel, Kere{\v s},
  Stern, {Faucher-Gigu{\`e}re}, Bullock, Wheeler, {El-Badry}, Loebman, Moreno,
  {Boylan-Kolchin}, \& Quataert}]{hopkinsWhatCausesFormation2023}
Hopkins, P.~F., Gurvich, A.~B., Shen, X., {et~al.} 2023, \bibinfo{title}{What
  Causes the Formation of Discs and End of Bursty Star Formation?} Monthly
  Notices of the Royal Astronomical Society, 525, 2241,
  \dodoi{10.1093/mnras/stad1902}

\bibitem[{P.~F. Hopkins {et~al.}(2024)Hopkins, Grudic, Su, Wellons,
  {Angles-Alcazar}, Steinwandel, Guszejnov, Murray, {Faucher-Giguere},
  Quataert, \& Keres}]{hopkinsFORGEdFIREResolving2024}
Hopkins, P.~F., Grudic, M.~Y., Su, K.-Y., {et~al.} 2024,
  \bibinfo{title}{{{FORGE}}'d in {{FIRE}}: {{Resolving}} the {{End}} of {{Star
  Formation}} and {{Structure}} of {{AGN Accretion Disks}} from {{Cosmological
  Initial Conditions}},} The Open Journal of Astrophysics, 7, 18,
  \dodoi{10.21105/astro.2309.13115}

\bibitem[{K. Inayoshi(2025)Inayoshi}]{inayoshiLittleRedDots2025}
Inayoshi, K. 2025, \bibinfo{title}{Little Red Dots as the Very First Activity
  of Black Hole Growth,} The Astrophysical Journal Letters, 988, L22,
  \dodoi{10.3847/2041-8213/adea66}

\bibitem[{K. Inayoshi {et~al.}(2026)Inayoshi, Murase, \&
  Kashiyama}]{inayoshiSpectralUniformityLittle2026}
Inayoshi, K., Murase, K., \& Kashiyama, K. 2026, \bibinfo{title}{Spectral
  {{Uniformity}} of {{Little Red Dots}}: {{A Natural Outcome}} of {{Coevolving
  Seed Black Holes}} and {{Nascent Starbursts}},} The Astrophysical Journal,
  1000, 90, \dodoi{10.3847/1538-4357/ae42ce}

\bibitem[{X. Ji {et~al.}(2025{\natexlab{a}})Ji, Maiolino, {\"U}bler, Scholtz,
  D'Eugenio, Sun, Perna, Turner, Carniani, Arribas, Bennett, Bunker, Charlot,
  Cresci, Curti, Egami, Fabian, Inayoshi, Isobe, Jones, Juod{\v z}balis,
  Kumari, Lyu, Mazzolari, Parlanti, Robertson, Rodr{\'i}guez Del~Pino,
  Schneider, Sijacki, Tacchella, Trinca, Valiante, Venturi, Volonteri, Willott,
  Witten, \& Witstok}]{jiBlackTHUNDERNonstellarBalmer2025}
Ji, X., Maiolino, R., {\"U}bler, H., {et~al.} 2025{\natexlab{a}},
  \bibinfo{title}{{{BlackTHUNDER}} - {{A}} Non-Stellar {{Balmer}} Break in a
  Black Hole-Dominated Little Red Dot at z = 7.04,} Monthly Notices of the
  Royal Astronomical Society, 544, 3900, \dodoi{10.1093/mnras/staf1867}

\bibitem[{X. Ji {et~al.}(2025{\natexlab{b}})Ji, Maiolino, Ferland, D'Eugenio,
  Bhatawdekar, Charlot, Chevallard, Curti, {Curtis-Lake}, Hainline, Ji,
  Robertson, Rodr{\'i}guez Del~Pino, Scholtz, Tacchella, Williams, \&
  Witstok}]{jiJADESSmallBlue2025}
Ji, X., Maiolino, R., Ferland, G., {et~al.} 2025{\natexlab{b}},
  \bibinfo{title}{{{JADES}} -- the Small Blue Bump in {{GN-z11}}: Insights into
  the Nuclear Region of a Galaxy at z = 10.6,} Monthly Notices of the Royal
  Astronomical Society, 541, 2134, \dodoi{10.1093/mnras/staf1083}

\bibitem[{F. Jiang {et~al.}(2021)Jiang, Dekel, Freundlich, {van den Bosch},
  Green, Hopkins, Benson, \& Du}]{jiangSatGenSemianalyticalSatellite2021}
Jiang, F., Dekel, A., Freundlich, J., {et~al.} 2021,
  \bibinfo{title}{{{SatGen}}: A Semi-Analytical Satellite Galaxy Generator -
  {{I}}. {{The}} Model and Its Application to {{Local-Group}} Satellite
  Statistics,} Monthly Notices of the Royal Astronomical Society, 502, 621,
  \dodoi{10.1093/mnras/staa4034}

\bibitem[{F. Jiang {et~al.}(2025)Jiang, Jia, Zheng, Ho, Inayoshi, Shen,
  Vogelsberger, \& Feng}]{jiangFormationLittleRed2025}
Jiang, F., Jia, Z., Zheng, H., {et~al.} 2025, Formation of the {{Little Red
  Dots}} from the {{Core-collapse}} of {{Self-interacting Dark Matter Halos}},
  arXiv, \dodoi{10.48550/arXiv.2503.23710}

\bibitem[{W. Jiang {et~al.}(2025)Jiang, Han, Dong, \&
  He}]{jiangSelfsimilarDecompositionHierarchical2025}
Jiang, W., Han, J., Dong, F., \& He, F. 2025, \bibinfo{title}{Self-Similar
  {{Decomposition}} of the {{Hierarchical Merger Tree}} of {{Dark Matter
  Halos}},} The Astrophysical Journal, 988, 160,
  \dodoi{10.3847/1538-4357/ade439}

\bibitem[{D. Kido {et~al.}(2025)Kido, Ioka, Hotokezaka, Inayoshi, \&
  Irwin}]{kidoBlackHoleEnvelopes2025}
Kido, D., Ioka, K., Hotokezaka, K., Inayoshi, K., \& Irwin, C.~M. 2025,
  \bibinfo{title}{Black Hole Envelopes in {{Little Red Dots}},} Monthly Notices
  of the Royal Astronomical Society, 544, 3407, \dodoi{10.1093/mnras/staf1898}

\bibitem[{V. Kokorev {et~al.}(2024)Kokorev, Caputi, Greene, Dayal, Trebitsch,
  Cutler, Fujimoto, Labb{\'e}, Miller, Iani, {Navarro-Carrera}, \&
  Rinaldi}]{kokorevCensusPhotometricallySelected2024}
Kokorev, V., Caputi, K.~I., Greene, J.~E., {et~al.} 2024, \bibinfo{title}{A
  Census of Photometrically Selected Little Red Dots at 4 \&lt; z \&lt; 9 in
  {{JWST}} Blank Fields,} The Astrophysical Journal, 968, 38,
  \dodoi{10.3847/1538-4357/ad4265}

\bibitem[{S. Koudmani {et~al.}(2024)Koudmani, Somerville, Sijacki, Bourne,
  Jiang, \& Profit}]{koudmaniUnifiedAccretionDisc2024}
Koudmani, S., Somerville, R.~S., Sijacki, D., {et~al.} 2024, \bibinfo{title}{A
  Unified Accretion Disc Model for Supermassive Black Holes in Galaxy Formation
  Simulations: Method and Implementation,} Monthly Notices of the Royal
  Astronomical Society, 532, 60, \dodoi{10.1093/mnras/stae1422}

\bibitem[{A. Kubota \& C. Done(2019)Kubota \&
  Done}]{kubotaModellingSpectralEnergy2019}
Kubota, A., \& Done, C. 2019, \bibinfo{title}{Modelling the Spectral Energy
  Distribution of Super-{{Eddington}} Quasars,} Monthly Notices of the Royal
  Astronomical Society, 489, 524, \dodoi{10.1093/mnras/stz2140}

\bibitem[{I. Labbe {et~al.}(2025)Labbe, Greene, Bezanson, Fujimoto, Furtak,
  Goulding, Matthee, Naidu, Oesch, Atek, Brammer, Chemerynska, Coe, Cutler,
  Dayal, Feldmann, Franx, Glazebrook, Leja, Maseda, Marchesini, Nanayakkara,
  Nelson, Pan, Papovich, Price, Suess, Wang, Weaver, Whitaker, Williams, \&
  Zitrin}]{labbeUNCOVERCandidateRed2025}
Labbe, I., Greene, J.~E., Bezanson, R., {et~al.} 2025,
  \bibinfo{title}{{{UNCOVER}}: {{Candidate Red Active Galactic Nuclei}} at 3
  {$<$} z {$<$} 7 with {{JWST}} and {{ALMA}},} The Astrophysical Journal, 978,
  92, \dodoi{10.3847/1538-4357/ad3551}

\bibitem[{E. Lambrides {et~al.}(2026)Lambrides, Larson, Garofali, Ptak,
  Chiaberge, Long, Hutchison, Norman, McKinney, Akins, Berg, Chisholm, Civano,
  Cloonan, Endsley, Faisst, Gilli, Gillman, Hirschmann, Kartaltepe, Kocevski,
  Kokorev, Pacucci, Richardson, Stiavelli, \&
  Whalen}]{lambridesCaseSuperEddingtonAccretion2026}
Lambrides, E., Larson, R.~L., Garofali, K., {et~al.} 2026, \bibinfo{title}{The
  Case for Super-{{Eddington}} Accretion in {{JWST}} Broad-Line Active Galactic
  Nuclei during the First Billion Years,} Nature Astronomy,
  \dodoi{10.1038/s41550-026-02813-w}

\bibitem[{M.~A. Latif {et~al.}(2022)Latif, Whalen, Khochfar, Herrington, \&
  Woods}]{latifTurbulentColdFlows2022}
Latif, M.~A., Whalen, D.~J., Khochfar, S., Herrington, N.~P., \& Woods, T.~E.
  2022, \bibinfo{title}{Turbulent Cold Flows Gave Birth to the First Quasars,}
  Nature, 607, 48, \dodoi{10.1038/s41586-022-04813-y}

\bibitem[{H. Li {et~al.}(2025)Li, Chen, Wang, \&
  Mo}]{liPhysicalProcessesCoevolution2025}
Li, H., Chen, Y., Wang, H., \& Mo, H. 2025, \bibinfo{title}{Physical Processes
  behind the Co-Evolution of Haloes, Galaxies, and Supermassive Black Holes in
  the {{IllustrisTNG}} Simulation,} Monthly Notices of the Royal Astronomical
  Society, 543, 1878, \dodoi{10.1093/mnras/staf1594}

\bibitem[{Z. Li {et~al.}(2025)Li, Inayoshi, Chen, Ichikawa, \&
  Ho}]{liLittleRedDots2025}
Li, Z., Inayoshi, K., Chen, K., Ichikawa, K., \& Ho, L.~C. 2025,
  \bibinfo{title}{Little {{Red Dots}}: {{Rapidly Growing Black Holes Reddened}}
  by {{Extended Dusty Flows}},} The Astrophysical Journal, 980, 36,
  \dodoi{10.3847/1538-4357/ada5fb}

\bibitem[{G.~V.
  Lipunova(1999)Lipunova}]{lipunovaSupercriticalDiskAccretion1999}
Lipunova, G.~V. 1999, \bibinfo{title}{Supercritical Disk Accretion with Mass
  Loss,} Astronomy Letters, 25, 508, \dodoi{10.48550/arXiv.astro-ph/9906324}

\bibitem[{H. Liu {et~al.}(2025)Liu, Jiang, Quataert, Greene, \&
  Ma}]{liuBalmerBreakOptical2025}
Liu, H., Jiang, Y.-F., Quataert, E., Greene, J.~E., \& Ma, Y. 2025,
  \bibinfo{title}{The {{Balmer Break}} and {{Optical Continuum}} of {{Little
  Red Dots}} from {{Super-Eddington Accretion}},} The Astrophysical Journal,
  994, 113, \dodoi{10.3847/1538-4357/ae0c19}

\bibitem[{Q. Ma {et~al.}(2026)Ma, Chen, \&
  Mo}]{maTwophaseFormationGalaxies2026}
Ma, Q., Chen, Y., \& Mo, H. 2026, \bibinfo{title}{Two-Phase Formation of
  Galaxies: The Coevolution between Galaxies and Dark Matter Haloes,} Monthly
  Notices of the Royal Astronomical Society, 548, stag562,
  \dodoi{10.1093/mnras/stag562}

\bibitem[{Y. Ma {et~al.}(2026)Ma, Greene, Setton, Goulding, Annunziatella, Fan,
  Kokorev, Labbe, Li, Lin, Marchesini, Matthee, Robbins, Sajina, Sawicki, \&
  Telford}]{maCountingLittleRed2026}
Ma, Y., Greene, J.~E., Setton, D.~J., {et~al.} 2026, \bibinfo{title}{Counting
  {{Little Red Dots}} at z \&lt; 4 with {{Ground-based Surveys}} and
  {{Spectroscopic Follow-up}},} The Astrophysical Journal, 1000, 59,
  \dodoi{10.3847/1538-4357/ae4596}

\bibitem[{R. Maiolino {et~al.}(2024)Maiolino, Scholtz, Witstok, Carniani,
  D'Eugenio, {de Graaff}, {\"U}bler, Tacchella, {Curtis-Lake}, Arribas, Bunker,
  Charlot, Chevallard, Curti, Looser, Maseda, Rawle, {Rodr{\'i}guez del Pino},
  Willott, Egami, Eisenstein, Hainline, Robertson, Williams, Willmer, Baker,
  Boyett, DeCoursey, Fabian, Helton, Ji, Jones, Kumari, Laporte, Nelson, Perna,
  Sandles, Shivaei, \& Sun}]{maiolinoSmallVigorousBlack2024}
Maiolino, R., Scholtz, J., Witstok, J., {et~al.} 2024, \bibinfo{title}{A Small
  and Vigorous Black Hole in the Early {{Universe}},} Nature, 627, 59,
  \dodoi{10.1038/s41586-024-07052-5}

\bibitem[{A. Marszewski {et~al.}(2026)Marszewski, {Faucher-Gigu{\`e}re}, Sun,
  {Angl{\'e}s-Alc{\'a}zar}, Feldmann, Su, Miller, \&
  Roy}]{marszewskiLittleRedDots2026}
Marszewski, A., {Faucher-Gigu{\`e}re}, C.-A., Sun, G., {et~al.} 2026,
  \bibinfo{title}{Little {{Red Dots}} on {{FIRE}}: {{The Ability}} of {{Bursty
  Galaxies}} to {{Host}} an {{Abundant Population}} of {{High-redshift AGN}},}
  The Astrophysical Journal, 1002, L13, \dodoi{10.3847/2041-8213/ae5d45}

\bibitem[{J. Matthee {et~al.}(2024)Matthee, Naidu, Brammer, Chisholm, Eilers,
  Goulding, Greene, Kashino, Labbe, Lilly, Mackenzie, Oesch, Weibel, Wuyts,
  Xiao, Bordoloi, Bouwens, {van Dokkum}, Illingworth, Kramarenko, Maseda,
  Mason, Meyer, Nelson, Reddy, Shivaei, Simcoe, \&
  Yue}]{mattheeLittleRedDots2024}
Matthee, J., Naidu, R.~P., Brammer, G., {et~al.} 2024, \bibinfo{title}{Little
  {{Red Dots}}: {{An Abundant Population}} of {{Faint Active Galactic Nuclei}}
  at z {$\sim$} 5 {{Revealed}} by the {{EIGER}} and {{FRESCO JWST Surveys}},}
  The Astrophysical Journal, 963, 129, \dodoi{10.3847/1538-4357/ad2345}

\bibitem[{H. Mo {et~al.}(2024)Mo, Chen, \& Wang}]{moTwophaseModelGalaxy2024}
Mo, H., Chen, Y., \& Wang, H. 2024, \bibinfo{title}{A Two-Phase Model of Galaxy
  Formation: {{I}}. {{The}} Growth of Galaxies and Supermassive Black Holes,}
  Monthly Notices of the Royal Astronomical Society, 532, 3808,
  \dodoi{10.1093/mnras/stae1727}

\bibitem[{R.~P. Naidu {et~al.}(2025)Naidu, Matthee, Katz, {de Graaff}, Oesch,
  Smith, Greene, Brammer, Weibel, Hviding, Chisholm, Labb{\textbackslash}'e,
  Simcoe, Witten, Atek, Baggen, Belli, Bezanson, Boogaard, Bose, {Covelo-Paz},
  Dayal, Fudamoto, Furtak, Giovinazzo, Goulding, Gronke, Heintz, Hirschmann,
  Illingworth, Inoue, Johnson, Leja, Leonova, McConachie, Maseda, Natarajan,
  Nelson, Setton, Shivaei, Sobral, Stefanon, Tacchella, Toft, Torralba, {van
  Dokkum}, {van der Wel}, Volonteri, Walter, Wang, \&
  Watson}]{naiduBlackHoleStar2025}
Naidu, R.~P., Matthee, J., Katz, H., {et~al.} 2025, A "{{Black Hole Star}}"
  {{Reveals}} the {{Remarkable Gas-Enshrouded Hearts}} of the {{Little Red
  Dots}}, arXiv, \dodoi{10.48550/arXiv.2503.16596}

\bibitem[{R.~P. Naidu {et~al.}(2026)Naidu, Oesch, Brammer, Weibel, Li, Matthee,
  Chisolm, Pollock, Heintz, Johnson, Shen, Hviding, Leja, Tacchella, Ganguly,
  Witten, Atek, Belli, Bose, Bouwens, Dayal, Decarli, {de Graaff}, Fudamoto,
  Giovinazzo, Greene, Illingworth, Inoue, Kane, Labbe, Leonova,
  {Marques-Chaves}, Meyer, Nelson, {Roberts-Borsani}, Schaerer, Simcoe,
  Stefanon, Sugahara, Toft, {van der Wel}, {van Dokkum}, Walter, Watson,
  Weaver, \& Whitaker}]{naiduCosmicMiracleRemarkably2026}
Naidu, R.~P., Oesch, P.~A., Brammer, G., {et~al.} 2026, \bibinfo{title}{A
  {{Cosmic Miracle}}: {{A Remarkably Luminous Galaxy}} at Zspec = 14.44
  {{Confirmed}} with {{JWST}},} The Open Journal of Astrophysics, 9, 56033,
  \dodoi{10.33232/001c.156033}

\bibitem[{D. Nelson {et~al.}(2019)Nelson, Springel, Pillepich,
  {Rodriguez-Gomez}, Torrey, Genel, Vogelsberger, Pakmor, Marinacci,
  Weinberger, Kelley, Lovell, Diemer, \&
  Hernquist}]{nelsonIllustrisTNGSimulationsPublic2019}
Nelson, D., Springel, V., Pillepich, A., {et~al.} 2019, \bibinfo{title}{The
  {{IllustrisTNG}} Simulations: Public Data Release,} Computational
  Astrophysics and Cosmology, 6, 2, \dodoi{10.1186/s40668-019-0028-x}

\bibitem[{P.~G. {P{\'e}rez-Gonz{\'a}lez}
  {et~al.}(2024){P{\'e}rez-Gonz{\'a}lez}, Barro, Rieke, Lyu, Rieke, Alberts,
  Williams, Hainline, Sun, Pusk{\'a}s, Annunziatella, Baker, Bunker, Egami, Ji,
  Johnson, Robertson, Rodr{\'i}guez Del~Pino, Rujopakarn, Shivaei, Tacchella,
  Willmer, \& Willott}]{perez-gonzalezWhatNatureLittle2024}
{P{\'e}rez-Gonz{\'a}lez}, P.~G., Barro, G., Rieke, G.~H., {et~al.} 2024,
  \bibinfo{title}{What {{Is}} the {{Nature}} of {{Little Red Dots}} and What
  {{Is Not}}, {{MIRI SMILES Edition}},} The Astrophysical Journal, 968, 4,
  \dodoi{10.3847/1538-4357/ad38bb}

\bibitem[{K. Perger {et~al.}(2025)Perger, Fogasy, Frey, \&
  Gab{\'a}nyi}]{pergerDeepSilenceRadio2025}
Perger, K., Fogasy, J., Frey, S., \& Gab{\'a}nyi, K.~{\'E}. 2025,
  \bibinfo{title}{Deep Silence: {{Radio}} Properties of Little Red Dots,}
  Astronomy and Astrophysics, 693, L2, \dodoi{10.1051/0004-6361/202452422}

\bibitem[{A. Pillepich {et~al.}(2018)Pillepich, Nelson, Hernquist, Springel,
  Pakmor, Torrey, Weinberger, Genel, Naiman, Marinacci, \&
  Vogelsberger}]{pillepichFirstResultsIllustrisTNG2018}
Pillepich, A., Nelson, D., Hernquist, L., {et~al.} 2018, \bibinfo{title}{First
  Results from the {{IllustrisTNG}} Simulations: The Stellar Mass Content of
  Groups and Clusters of Galaxies,} Monthly Notices of the Royal Astronomical
  Society, 475, 648, \dodoi{10.1093/mnras/stx3112}

\bibitem[{ {Planck Collaboration} {et~al.}(2016){Planck Collaboration}, Ade,
  Aghanim, Arnaud, Ashdown, Aumont, Baccigalupi, Banday, Barreiro, Bartlett,
  Bartolo, Battaner, Battye, Benabed, Beno{\^i}t, {Benoit-L{\'e}vy}, Bernard,
  Bersanelli, Bielewicz, Bock, Bonaldi, Bonavera, Bond, Borrill, Bouchet,
  Boulanger, Bucher, Burigana, Butler, Calabrese, Cardoso, Catalano, Challinor,
  Chamballu, Chary, Chiang, Chluba, Christensen, Church, Clements, Colombi,
  Colombo, Combet, Coulais, Crill, Curto, Cuttaia, Danese, Davies, Davis, {de
  Bernardis}, {de Rosa}, {de Zotti}, Delabrouille, D{\'e}sert, Di~Valentino,
  Dickinson, Diego, Dolag, Dole, Donzelli, Dor{\'e}, Douspis, Ducout, Dunkley,
  Dupac, Efstathiou, Elsner, En{\ss}lin, Eriksen, Farhang, Fergusson, Finelli,
  Forni, Frailis, Fraisse, Franceschi, Frejsel, Galeotta, Galli, Ganga,
  Gauthier, Gerbino, Ghosh, Giard, {Giraud-H{\'e}raud}, Giusarma, Gjerl{\o}w,
  {Gonz{\'a}lez-Nuevo}, G{\'o}rski, Gratton, Gregorio, Gruppuso, Gudmundsson,
  Hamann, Hansen, Hanson, Harrison, Helou, {Henrot-Versill{\'e}},
  {Hern{\'a}ndez-Monteagudo}, Herranz, Hildebrandt, Hivon, Hobson, Holmes,
  Hornstrup, Hovest, Huang, Huffenberger, Hurier, Jaffe, Jaffe, Jones, Juvela,
  Keih{\"a}nen, Keskitalo, Kisner, Kneissl, Knoche, Knox, Kunz, {Kurki-Suonio},
  Lagache, L{\"a}hteenm{\"a}ki, Lamarre, Lasenby, Lattanzi, Lawrence, Leahy,
  Leonardi, Lesgourgues, Levrier, Lewis, Liguori, Lilje, {Linden-V{\o}rnle},
  {L{\'o}pez-Caniego}, Lubin, {Mac{\'i}as-P{\'e}rez}, Maggio, Maino, Mandolesi,
  Mangilli, Marchini, Maris, Martin, Martinelli, {Mart{\'i}nez-Gonz{\'a}lez},
  Masi, Matarrese, McGehee, Meinhold, Melchiorri, Melin, Mendes, Mennella,
  Migliaccio, Millea, Mitra, {Miville-Desch{\^e}nes}, Moneti, Montier,
  Morgante, Mortlock, Moss, Munshi, Murphy, Naselsky, Nati, Natoli,
  Netterfield, {N{\o}rgaard-Nielsen}, Noviello, Novikov, Novikov, Oxborrow,
  Paci, Pagano, Pajot, Paladini, Paoletti, Partridge, Pasian, Patanchon,
  Pearson, Perdereau, Perotto, Perrotta, Pettorino, Piacentini, Piat,
  Pierpaoli, Pietrobon, Plaszczynski, Pointecouteau, Polenta, Popa, Pratt,
  Pr{\'e}zeau, Prunet, Puget, Rachen, Reach, Rebolo, Reinecke, Remazeilles,
  Renault, Renzi, Ristorcelli, Rocha, Rosset, Rossetti, Roudier, {Rouill{\'e}
  d'Orfeuil}, {Rowan-Robinson}, {Rubi{\~n}o-Mart{\'i}n}, Rusholme, Said,
  Salvatelli, Salvati, Sandri, Santos, Savelainen, Savini, Scott, Seiffert,
  Serra, Shellard, Spencer, Spinelli, Stolyarov, Stompor, Sudiwala, Sunyaev,
  Sutton, {Suur-Uski}, Sygnet, Tauber, Terenzi, Toffolatti, Tomasi, Tristram,
  Trombetti, Tucci, Tuovinen, T{\"u}rler, Umana, Valenziano, Valiviita,
  Van~Tent, Vielva, Villa, Wade, Wandelt, Wehus, White, White, Wilkinson, Yvon,
  Zacchei, \& Zonca}]{planckcollaborationPlanck2015Results2016}
{Planck Collaboration}, Ade, P. A.~R., Aghanim, N., {et~al.} 2016,
  \bibinfo{title}{Planck 2015 Results. {{XIII}}. {{Cosmological}} Parameters,}
  Astronomy and Astrophysics, 594, A13, \dodoi{10.1051/0004-6361/201525830}

\bibitem[{M.~G. Roberts {et~al.}(2026)Roberts, Braff, Garg, Profumo, \&
  Jeltema}]{robertsLittleRedDots2026}
Roberts, M.~G., Braff, L., Garg, A., Profumo, S., \& Jeltema, T. 2026,
  \bibinfo{title}{Little {{Red Dots}} from Ultra-Strongly Self-Interacting Dark
  Matter,} Journal of Cosmology and Astroparticle Physics, 2026, 003,
  \dodoi{10.1088/1475-7516/2026/05/003}

\bibitem[{T.~P. Robitaille {et~al.}(2013)Robitaille, Tollerud, Greenfield,
  Droettboom, Bray, Aldcroft, Davis, Ginsburg, {Price-Whelan}, Kerzendorf,
  Conley, Crighton, Barbary, Muna, Ferguson, Grollier, Parikh, Nair,
  G{\"u}nther, Deil, Woillez, Conseil, Kramer, Turner, Singer, Fox, Weaver,
  Zabalza, Edwards, Bostroem, Burke, Casey, Crawford, Dencheva, Ely, Jenness,
  Labrie, Lim, Pierfederici, Pontzen, Ptak, Refsdal, Servillat, \&
  Streicher}]{robitailleAstropyCommunityPython2013}
Robitaille, T.~P., Tollerud, E.~J., Greenfield, P., {et~al.} 2013,
  \bibinfo{title}{Astropy: {{A}} Community {{Python}} Package for Astronomy,}
  Astronomy \& Astrophysics, 558, A33, \dodoi{10.1051/0004-6361/201322068}

\bibitem[{X. Shen {et~al.}(2023)Shen, Vogelsberger, {Boylan-Kolchin},
  Tacchella, \& Kannan}]{shenImpactUVVariability2023}
Shen, X., Vogelsberger, M., {Boylan-Kolchin}, M., Tacchella, S., \& Kannan, R.
  2023, \bibinfo{title}{The Impact of {{UV}} Variability on the Abundance of
  Bright Galaxies at z {$\geq$} 9,} Monthly Notices of the Royal Astronomical
  Society, 525, 3254, \dodoi{10.1093/mnras/stad2508}

\bibitem[{X. Shen {et~al.}(2026)Shen, Zier, Smith, Liu, Kannan, Bulichi,
  Koehler, Springel, Vogelsberger, Hernquist, Naidu, {de Graaff}, Pizzati,
  Alexander, Ho, Kokorev, Leung, Eilers, \&
  Hickox}]{shenLuminaProjectDemographics2026}
Shen, X., Zier, O., Smith, A., {et~al.} 2026, The {{Lumina Project}}: {{The
  Demographics}} of {{Active Galactic Nuclei}} from {{Quasars}} to {{Little Red
  Dots}} at \$z\textbackslash geq 3\$, arXiv, \dodoi{10.48550/arXiv.2605.24112}

\bibitem[{D. Sijacki {et~al.}(2007)Sijacki, Springel, Di~Matteo, \&
  Hernquist}]{sijackiUnifiedModelAGN2007}
Sijacki, D., Springel, V., Di~Matteo, T., \& Hernquist, L. 2007,
  \bibinfo{title}{A Unified Model for {{AGN}} Feedback in Cosmological
  Simulations of Structure Formation,} Monthly Notices of the Royal
  Astronomical Society, 380, 877, \dodoi{10.1111/j.1365-2966.2007.12153.x}

\bibitem[{A. Sivasankaran {et~al.}(2025)Sivasankaran, Blecha, Torrey, Kelley,
  Bhowmick, Vogelsberger, Hernquist, Marinacci, \&
  Sales}]{sivasankaranAGNFeedbackIsolated2025}
Sivasankaran, A., Blecha, L., Torrey, P., {et~al.} 2025,
  \bibinfo{title}{{{AGN}} Feedback in Isolated Galaxies with a {{SMUGGLE}}
  Multiphase {{ISM}},} Monthly Notices of the Royal Astronomical Society, 537,
  817, \dodoi{10.1093/mnras/staf062}

\bibitem[{{\'A}. Sk{\'u}lad{\'o}ttir {et~al.}(2024)Sk{\'u}lad{\'o}ttir,
  Koutsouridou, Vanni, Amarsi, Lucchesi, Salvadori, \&
  Aguado}]{skuladottirPairinstabilitySupernovaOrigin2024}
Sk{\'u}lad{\'o}ttir, {\'A}., Koutsouridou, I., Vanni, I., {et~al.} 2024,
  \bibinfo{title}{On the {{Pair-instability Supernova Origin}} of
  {{J1010}}+2358,} The Astrophysical Journal, 968, L23,
  \dodoi{10.3847/2041-8213/ad4b1a}

\bibitem[{R.~S. Somerville {et~al.}(2008)Somerville, Hopkins, Cox, Robertson,
  \& Hernquist}]{somervilleSemianalyticModelCoevolution2008}
Somerville, R.~S., Hopkins, P.~F., Cox, T.~J., Robertson, B.~E., \& Hernquist,
  L. 2008, \bibinfo{title}{A Semi-Analytic Model for the Co-Evolution of
  Galaxies, Black Holes and Active Galactic Nuclei,} Monthly Notices of the
  Royal Astronomical Society, 391, 481,
  \dodoi{10.1111/j.1365-2966.2008.13805.x}

\bibitem[{D. Spinoso {et~al.}(2023)Spinoso, Bonoli, Valiante, Schneider, \&
  {Izquierdo-Villalba}}]{spinosoMultiflavourSMBHSeeding2023}
Spinoso, D., Bonoli, S., Valiante, R., Schneider, R., \& {Izquierdo-Villalba},
  D. 2023, \bibinfo{title}{Multiflavour {{SMBH}} Seeding and Evolution in
  Cosmological Environments,} Monthly Notices of the Royal Astronomical
  Society, 518, 4672, \dodoi{10.1093/mnras/stac3169}

\bibitem[{J. Stern {et~al.}(2021)Stern, {Faucher-Gigu{\`e}re}, Fielding,
  Quataert, Hafen, Gurvich, Ma, Byrne, {El-Badry}, {Angl{\'e}s-Alc{\'a}zar},
  Chan, Feldmann, Kere{\v s}, Wetzel, Murray, \&
  Hopkins}]{sternVirializationInnerCGM2021}
Stern, J., {Faucher-Gigu{\`e}re}, C.-A., Fielding, D., {et~al.} 2021,
  \bibinfo{title}{Virialization of the {{Inner CGM}} in the {{FIRE
  Simulations}} and {{Implications}} for {{Galaxy Disks}}, {{Star Formation}},
  and {{Feedback}},} The Astrophysical Journal, 911, 88,
  \dodoi{10.3847/1538-4357/abd776}

\bibitem[{K.-Y. Su {et~al.}(2023)Su, Bryan, Haiman, Somerville, Hayward, \&
  {Faucher-Gigu{\`e}re}}]{suSelfregulationBlackHole2023}
Su, K.-Y., Bryan, G.~L., Haiman, Z., {et~al.} 2023,
  \bibinfo{title}{Self-Regulation of Black Hole Accretion via Jets in Early
  Protogalaxies,} Monthly Notices of the Royal Astronomical Society, 520, 4258,
  \dodoi{10.1093/mnras/stad252}

\bibitem[{S. Tacchella {et~al.}(2023)Tacchella, Eisenstein, Hainline, Johnson,
  Baker, Helton, Robertson, Suess, Chen, Nelson, Pusk{\'a}s, Sun, Alberts,
  Egami, Hausen, Rieke, Rieke, Shivaei, Williams, Willmer, Bunker, Cameron,
  Carniani, Charlot, Curti, {Curtis-Lake}, Looser, Maiolino, Maseda, Rawle,
  Rix, Smit, {\"U}bler, Willott, Witstok, Baum, Bhatawdekar, Boyett, Danhaive,
  {de Graaff}, Endsley, Ji, Lyu, Sandles, Saxena, Scholtz, Topping, \&
  Whitler}]{tacchellaJADESImagingGNz112023}
Tacchella, S., Eisenstein, D.~J., Hainline, K., {et~al.} 2023,
  \bibinfo{title}{{{JADES Imaging}} of {{GN-z11}}: {{Revealing}} the
  {{Morphology}} and {{Environment}} of a {{Luminous Galaxy}} 430 {{Myr}} after
  the {{Big Bang}},} The Astrophysical Journal, 952, 74,
  \dodoi{10.3847/1538-4357/acdbc6}

\bibitem[{A. Trinca {et~al.}(2026)Trinca, Lupi, Haardt, \&
  Madau}]{trincaYouCantSee2026}
Trinca, A., Lupi, A., Haardt, F., \& Madau, P. 2026, \bibinfo{title}{You Can't
  See Me: {{Super-Eddington}} Growth Hindering {{X-ray}} Detection in High-z
  Broad-Line Active Galactic Nuclei,} Astronomy and Astrophysics, 710, A289,
  \dodoi{10.1051/0004-6361/202659544}

\bibitem[{K. Wang {et~al.}(2023)Wang, Peng, Liu, Mihos, C{\^o}t{\'e},
  Ferrarese, Taylor, Blakeslee, Cuillandre, Duc, Guhathakurta, Gwyn, Ko, Lan{\c
  c}on, Lim, MacArthur, Puzia, Roediger, Sales, {S{\'a}nchez-Janssen},
  Spengler, Toloba, Zhang, \& Zhu}]{wangEvolutionaryContinuumNucleated2023}
Wang, K., Peng, E.~W., Liu, C., {et~al.} 2023, \bibinfo{title}{An Evolutionary
  Continuum from Nucleated Dwarf Galaxies to Star Clusters,} Nature, 623, 296,
  \dodoi{10.1038/s41586-023-06650-z}

\bibitem[{Z. Wang {et~al.}(2026)Wang, Jiang, Zheng, Shen, Jia, Ho, Inayoshi, \&
  Jiang}]{wangHaloAssemblyBias2026}
Wang, Z., Jiang, F., Zheng, H., {et~al.} 2026, Halo Assembly Bias in the Early
  {{Universe}}: A Clustering Probe of the Origin of the {{Little Red Dots}},
  arXiv, \dodoi{10.48550/arXiv.2603.15736}

\bibitem[{R. Weinberger {et~al.}(2017)Weinberger, Springel, Hernquist,
  Pillepich, Marinacci, Pakmor, Nelson, Genel, Vogelsberger, Naiman, \&
  Torrey}]{weinbergerSimulatingGalaxyFormation2017}
Weinberger, R., Springel, V., Hernquist, L., {et~al.} 2017,
  \bibinfo{title}{Simulating Galaxy Formation with Black Hole Driven Thermal
  and Kinetic Feedback,} Monthly Notices of the Royal Astronomical Society,
  465, 3291, \dodoi{10.1093/mnras/stw2944}

\bibitem[{J.~H. Wise {et~al.}(2019)Wise, Regan, O'Shea, Norman, Downes, \&
  Xu}]{wiseFormationMassiveBlack2019}
Wise, J.~H., Regan, J.~A., O'Shea, B.~W., {et~al.} 2019,
  \bibinfo{title}{Formation of Massive Black Holes in Rapidly Growing
  Pre-Galactic Gas Clouds,} Nature, 566, 85, \dodoi{10.1038/s41586-019-0873-4}

\bibitem[{Q.-F. Xing {et~al.}(2023)Xing, Zhao, Liu, Heger, Han, Aoki, Chen,
  Ishigaki, Li, \& Zhao}]{xingMetalpoorStarAbundances2023}
Xing, Q.-F., Zhao, G., Liu, Z.-W., {et~al.} 2023, \bibinfo{title}{A Metal-Poor
  Star with Abundances from a Pair-Instability Supernova,} Nature, 618, 712,
  \dodoi{10.1038/s41586-023-06028-1}

\bibitem[{F. Yuan \& R. Narayan(2014)Yuan \&
  Narayan}]{yuanHotAccretionFlows2014}
Yuan, F., \& Narayan, R. 2014, \bibinfo{title}{Hot {{Accretion Flows Around
  Black Holes}},} Annual Review of Astronomy and Astrophysics, 52, 529,
  \dodoi{10.1146/annurev-astro-082812-141003}

\bibitem[{M. Yue {et~al.}(2024)Yue, Eilers, Ananna, Panagiotou, Kara, \&
  Miyaji}]{yueStackingXRayObservations2024}
Yue, M., Eilers, A.-C., Ananna, T.~T., {et~al.} 2024, \bibinfo{title}{Stacking
  {{X-Ray Observations}} of "{{Little Red Dots}}": {{Implications}} for {{Their
  Active Galactic Nucleus Properties}},} The Astrophysical Journal, 974, L26,
  \dodoi{10.3847/2041-8213/ad7eba}

\bibitem[{Y. Zhang {et~al.}(2025)Zhang, Ding, Yang, Lambrides, Akins, Battisti,
  Casey, Chen, Cox, Faisst, Franco, Haghjoo, Ho, Inayoshi, Jin, Karmen,
  Koekemoer, Kartaltepe, Liao, Gozaliasl, Onoue, Kokorev, Roy, Rich, Silverman,
  Tanaka, You, Yesuf, \& Zavala}]{zhangUnveilingExtendedComponents2025}
Zhang, Y., Ding, X., Yang, L., {et~al.} 2025, Unveiling {{Extended Components}}
  of '{{Little Red Dots}}' in {{Rest-Frame Optical}}, arXiv,
  \dodoi{10.48550/arXiv.2510.25830}

\bibitem[{Z. Zhang {et~al.}(2025)Zhang, Chen, Rong, Wang, Mo, Luo, \&
  Li}]{zhangUnexpectedClusteringPattern2025}
Zhang, Z., Chen, Y., Rong, Y., {et~al.} 2025, \bibinfo{title}{Unexpected
  Clustering Pattern in Dwarf Galaxies Challenges Formation Models,} Nature,
  642, 47, \dodoi{10.1038/s41586-025-08965-5}

\end{thebibliography}
\bibliographystyle{aasjournalv7}



\end{document}